\RequirePackage{rotating}
\documentclass[twocolumn]{aastex631}

\usepackage{epstopdf}
\usepackage{rotating}
\usepackage{newtxtext,newtxmath}

\usepackage[T1]{fontenc}

\usepackage{graphicx}	
\usepackage{amsmath}	
\usepackage{amssymb}
\usepackage{epstopdf}
\usepackage{color}
\usepackage[caption = false]{subfig}

\shorttitle{Visual Binaries in Gaia eDR3 with 2MASS}
\shortauthors{Medan \& L\'{e}pine}

\begin{document}
	
	\title{Detecting New Visual Binaries in Gaia DR3 with Gaia and 2MASS Photometry I. New Candidate Binaries Within 200 pc of the Sun\footnote{Accepted 2023 October 02. Received 2023 October 02; in original form 2023 June 26}}

\author[0000-0003-3410-5794]{Ilija Medan}
\affiliation{Department of Physics and Astronomy,
	Georgia State University,
	Atlanta, GA 30302, USA}
\affiliation{Department of Physics and Astronomy,
	Vanderbilt University,
	Nashville, TN 37235, USA}

\author[0000-0002-2437-2947]{S\'{e}bastien L\'{e}pine}
\affiliation{Department of Physics and Astronomy,
	Georgia State University,
	Atlanta, GA 30302, USA}


\begin{abstract}
We present a method to identify likely visual binaries in Gaia eDR3 that does not rely on parallax or proper motion. This method utilizes the various PSF sizes of 2MASS/Gaia, where at $<2.5$" two stars may be unresolved in 2MASS but resolved by Gaia. Due to this, if close neighbors listed in Gaia are a resolved pair, the associated 2MASS source will have a predictable excess in the J-band that depends on the $\Delta G$ of the pair. We demonstrate that the expected relationship between 2MASS excess and $\Delta G$ differs for chance alignments, as compared to true binary systems, when parameters like magnitude and location on the sky are also considered. Using these multidimensional distributions, we compute the likelihood of a close pair of stars to be a chance alignment, resulting in \textcolor{black}{a total(clean) catalog of 68,725(50,230) likely binaries within 200 pc with a completeness rate of $\sim75\%$($\sim64\%$) and contamination rate of $\sim14\%$($\sim0.4\%$).} Within this, we find \textcolor{black}{590} previously unidentified binaries from Gaia eDR3 with projected physical separations $<30$ AU, where 138 systems were previously identified, and for $s<10$ AU we find that \textcolor{black}{4} out of \textcolor{black}{15} new \textcolor{black}{likely} binaries have not yet been observed with high-resolution imaging. We also demonstrate the potential of our catalog to determine physical separation distributions and binary fraction estimates, from this increase in low separation binaries. Overall, this catalog provides a good complement for the study of local binary populations by probing smaller physical separations and mass ratios, and provides prime targets for speckle monitoring.
\end{abstract}

\section{Introduction}

Gaia eDR3 provides a significant improvement in parallax and proper motion precision over DR2 \citep{gaiaedr3_summary}. It has however been noticed that eDR3 lists additional sources at small angular separations ($<0.4$") from many stars. These close neighbors, which were not present in DR2, are in excess of the expected distribution from random field star alignments \citep{gaiaedr3_catalog_valid}. It is difficult to determine if these close neighbors are spurious entries in the catalog (e.g. duplicates), or whether they are true detections of resolved sources, especially given that 74\% of neighbors at separations $<0.4$" only have a 2-parameter (position only) solution \citep{gaiaedr3_catalog_valid}. Even at slightly larger separations, fainter neighbors typically only have a 5-parameter (position, proper motion and parallax) solution if the separation is $>2$" \citep{gaiaedr3_astro_sol}.

On the other hand, one would expect Gaia to resolve physical companions at such angular separations, notably for relatively nearby stars. For solar-type stars, binary separations are expected to peak at $\sim45$ AU \citep{raghavan2010}, which corresponds to an angular separation of $\sim0.45$" at a distance of 100 pc. This implies that the push in resolution limit down to 0.4" could allow for $\sim55\%$ of solar-type binaries within 100 pc to be resolved. For the more common M dwarfs, binary separations are expected to peak at $\sim6$ AU \citep{Ward-Duong2015}, which corresponds to a much smaller angular separation of $\sim0.06$" at a distance of 100 pc. Based on the separation distribution from  \citet{Ward-Duong2015}, we would expect $\sim30\%$ of M dwarf binaries within 100 pc to be resolved by Gaia assuming a resolution limit of 0.4". For separations $>2$", where we expect most faint neighbors to have 5-parameter solutions, we would expect a much smaller  $\sim37\%$ of solar-type binaries and $\sim15\%$ of M dwarf binaries to be resolved within 100 pc. \textcolor{black}{We do note that improvements could be even better in future data releases than the numbers quoted above. While, nominally, equal brightness double stars could be detected at 0.23 arcsec in the along-scan and 0.70 arcsec in the across-scan direction \citep{Bruijne2015}, by \textit{Gaia} DR5 it is expected that with on-ground processing the full resolution of the instrument will allow for stars with separations of $\sim0.1$" to be resolved. With such a resolution, a large portion of local binaries will be detected, but as described above it is unclear how they can be effectively identified as 1) part of the local population and 2) as true binaries.}

\textcolor{black}{There are some additional parameters in the \textit{Gaia} DR3 data products that are useful in identifying possible binaries that could help push the above current boundaries. Primarily, the RUWE value has been used  to identify astrometric solutions that deviate significantly from a single star solution \citep{Belokurov2020}. A more powerful diagnostic for barely resolved systems, which applies to many of these very close separation systems, could be the \texttt{ipd\_frac\_multi\_peak} parameter. This parameter gives the percent of detections of more than one peak in the raw windows used for the astrometric processing for the source. Recently, \cite{Tokovinin2023} used \texttt{ipd\_frac\_multi\_peak} to select close pairs from the Gaia Catalog of Nearby Stars \citep[GCNS;][]{gcns} within 100 pc for follow up speckle observations. Out of the 1243 candidates observed, 506 inner pairs were resolved. This does demonstrate the usefulness of this parameter in detecting binaries, but it also shows that such double transits do not always correlate to true binary systems. An additional issue when trying to detect \textit{local} binaries is the presence of spurious solutions, which can occur for close source pairs depending on the scan angle for the observation, and will result in meaningless parallax and proper motion values \citep{gaiaedr3_catalog_valid}.}

One question therefore is whether we can confirm if these neighbors are actual companions rather than chance alignments of unrelated background sources, or even spurious entries, without the use of a 5-parameter astrometric solution. In this paper, we propose a photometric/statistical method to calculate \textcolor{black}{a quasi-}likelihood that an eDR3 neighbor to a nearby ($d<200$ pc) star is a true resolved companion. The method combines Gaia photometry of the primary stars and its alleged secondary with infrared magnitudes of the unresolved object measured by 2MASS, and estimates whether the measurements are consistent with the presence of a stellar companion. \textcolor{black}{We then combine these infrared excesses with a number of other measurements to detect likely binaries at small angular separations without the need for an astrometric solution for the secondary.}

In Section \ref{sec:data4}, we provide an overview of the datasets used in this study, which include a subset of Gaia eDR3 stars within 200 pc, subsets of likely chance alignments, and a list of expected photometric values based on stellar mass of main sequence stars. In Section \ref{sec:results4}, we compare the multidimensional distributions for the candidates and likely chance alignments, and use these to determine the likelihood that a pair of stars in the 200 pc sample is a chance alignment. In Section \ref{sec:discuss4}, we compare the identified likely binaries to visual binaries previously identified in Gaia eDR3. \textcolor{black}{We also} identify which stars in the sample have previously been observed with high resolution imaging techniques. Additionally, we demonstrate how the catalog has the potential to improve the estimate of the binary fraction of low-mass stars in the Solar Neighborhood. Finally, in Section \ref{sec:conclusions} we provide a summary of our results and a brief note about future observations that will be crucial to confirm the shortest separation binaries we have identified.

\section{Data}\label{sec:data4}

To examine the validity of the low-separation neighbors in Gaia eDR3, we combine 2MASS photometry with Gaia eDR3 data. This is useful because 2MASS has an angular resolution of $\sim 2$", which means that for all of these low-angular separation neighbors listed in eDR3, we can expect that if these are indeed two real sources, they will be unresolved in 2MASS and have photometric measurements consistent with a blended system.  In the case where a true detection is confirmed, we can further evaluate whether the resolved object is consistent with a physical companion, or whether it is more likely to be a chance alignment of an unrelated field star.

For true detection cases, we will be focusing on the subset of eDR3 stars with parallaxes placing them within 200 pc of the Sun. We further restrict our analysis to stars that have a 2MASS counterpart, determined from the  \texttt{gaiaedr3.tmass\_psc\_xsc\_best\_neighbour} entry in the Gaia archive. Additionally, we extract from eDR3 all stars listed to be within 2.5" of each 200 pc star to assemble our main sample of low-separation neighbors. Our initial subset comprises a total of 2,242,081 stars, of which 1,925,842 stars have no neighbor within 2.5" and 316,239 stars have at least one neighbor within 2.5". Additionally, for this analysis we do not use any star with $BP>20$ due to issues with the $BP$ and $RP$ magnitude measurements at the faint end in Gaia eDR3 \citep{gaiaedr3_catalog_valid}. This leaves 202,059 within 200 pc stars with at least one neighbor within 2.5" to use for the subsequent analysis. We note that 85,039 of these 200 pc stars have neighbors with no parallax measurements, which are the primary motivation for this study as they cannot be verified to be physical companions based on eDR3 astrometry alone.

To help determine whether the low-separation neighbor is a possible binary companion or a chance alignment field star, we also assemble a sample of field stars. These stars are selected directly from the 200 pc \textcolor{black}{sample} with at least one neighbor within 2.5" among those for which we already have a high level of confidence that they are chance alignments. These stars come from two populations; neighbors with parallax measurements and neighbors without parallax measurements. For the former, we select pairs of stars where the parallax measurements are significantly different, indicating the stars are not at the same distance and thus represent chance alignments:
\begin{equation}
	\frac{|\pi_1 - \pi_2|}{ \sqrt{\sigma_{\pi_1}^2 + \sigma_{\pi_2}^2} } > 6
\end{equation}
From the above condition, we identify pairs with parallaxes differences greater than $6\sigma$ to be chance alignments, as it has been shown that the parallax uncertainties are significantly underestimated for pairs with $\theta < 4$ arcseconds \citep{elbadry_2021}.

\begin{figure*}[!t]
	\centering
	\includegraphics[width=1\textwidth]{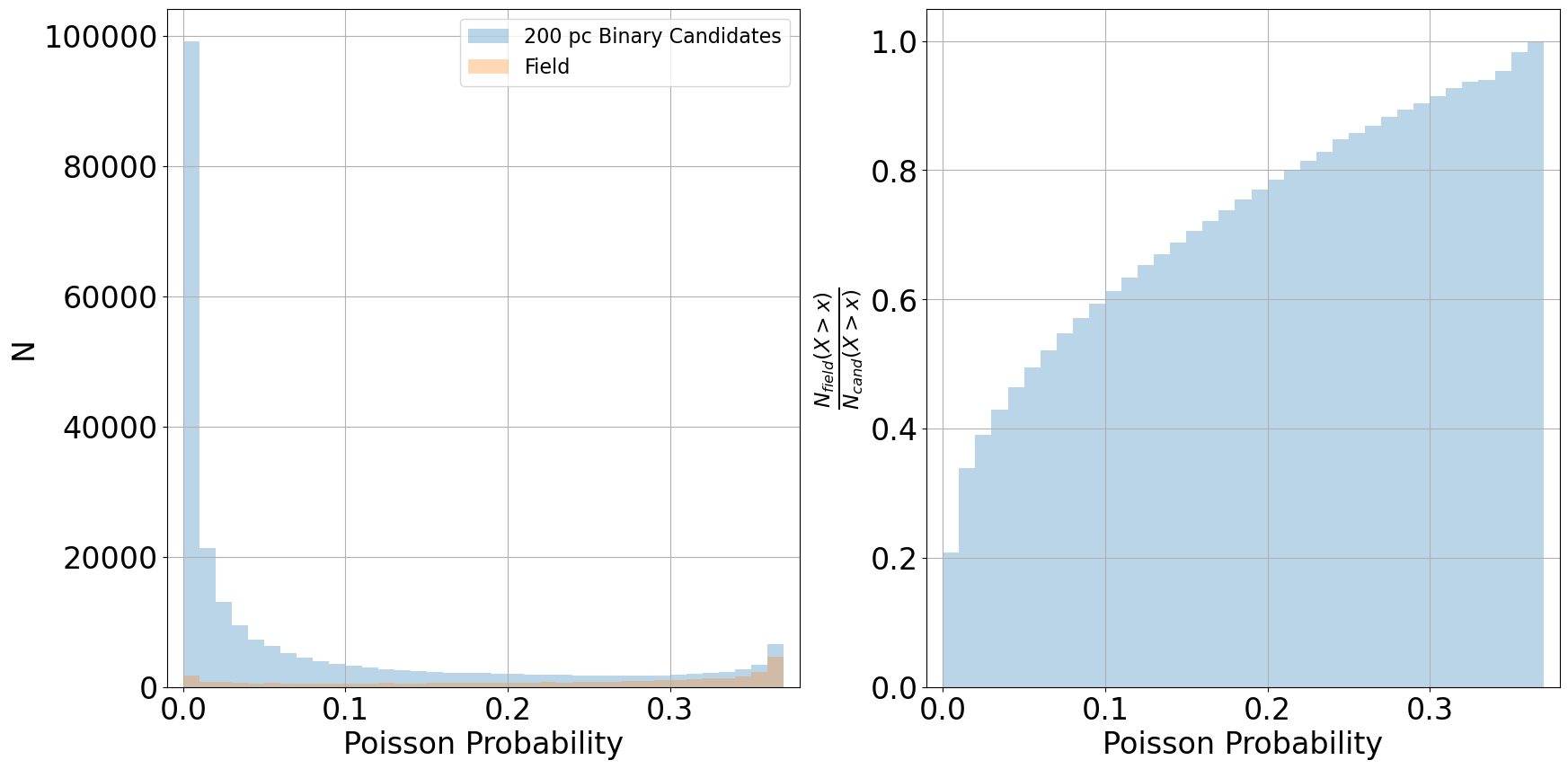}
	\caption{\textbf{Left Panel:} Distribution of Poisson probabilities for a companion to be a chance alignment based on the average local stellar field density (eqs. \ref{eq:local_dens}-\ref{eq:poss_prob}) for the 200 pc stars with a close neighbor (blue bins) and the surrounding field stars within 1 arcminute (orange bins). \textbf{Right Panel:} Ratio of the number of field stars to 200 pc stars with a close neighbor with probability greater than $x$. The ratio has been normalized such that it is assumed that 100\% of stars in the highest probability bin are field stars.\label{fig:poiss_probs}}
\end{figure*}

The above only provides a sample of chance alignments for stars with parallax measurements, which may have a different distributions in parameter space (e.g. magnitude) than chance alignments with no parallax measurements. To build a sample of highly likely chance alignments with no parallax measurements, we identify pairs in such high density fields that the companion is most likely to be a chance alignment. We do this as follows, first we query a region of 1 arcminute around every star and count the number of stars with $G$ magnitudes less than the fainter neighbor to the 200 pc source. This number then gives the expected average local stellar density:
\begin{equation}\label{eq:local_dens}
	\rho_{local} = \frac{N_{G < G_{ne}}}{\pi r^2}
\end{equation}
Then, given the average local stellar density ($\rho_{local}$), the expected number of stars at the distance between the 200 pc star and the close neighbor ($r_{ne}$) is $N_{ne} = \rho_{local} \times \pi r_{ne}^2$. Expecting then that the stellar density follows a Poisson distribution, the probability of finding exactly one star only up to a separation of $r_{ne}$ is:
\begin{equation}\label{eq:poss_prob}
	p(N=1) = N_{ne} e^{-N_{ne}}
\end{equation}
We calculate this Poisson probability for all the close neighbors of the 200 pc stars, and for all the stars in the 1 arcminute region around these stars (i.e. the ``field" stars). The resulting Poisson probabilities for the close neighbors (blue bins) and the nearby field stars (orange bins) are shown in the left panel of Figure \ref{fig:poiss_probs}. Here we see that the vast majority of the close neighbors to the 200 pc stars have low Poisson probabilities, indicating that (based on the local stellar density) it is very unlikely that the near neighbor is there by chance and is thus more consistent being in a physical companion. We also see that there are some close neighbors with high ($p(N=1)  > 0.3$) Poisson probabilities; these are in high density fields and are much more likely to be chance alignments. Additionally, these later, high probability stars have a distribution of Poisson probabilities that is similar to the distribution observed for the field stars. So, if we then assume that 100\% of the stars in the highest probability bin are field stars, we can normalize the probability distribution for field stars and estimate the fraction of 200 pc stars whose neighbor is likely a field star for Poisson probabilities greater than some value (right panel of Figure \ref{fig:poiss_probs}). This demonstrates that by selecting 200 pc stars with a neighbor having a probability $> 0.3$, that we should expect $\sim91\%$ of these neighbors to be field stars. In raw numbers this equates to 20,923 pairs, of which 9,678 have neighbors with no parallax measurements. The combination of this with the former sample will be used as the sample of highly likely chance alignments for the subsequent analysis.

\begin{figure*}[!t]
	\centering
	\includegraphics[width=1\textwidth]{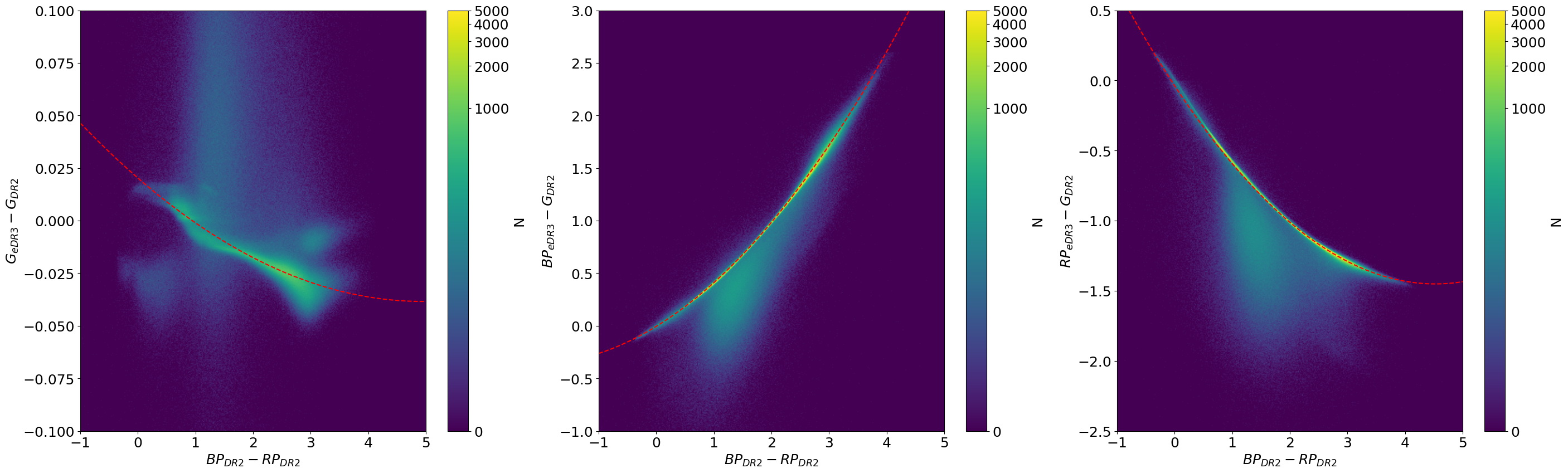}
	\caption{Relations between Gaia eDR3 and DR2 photometry for each Gaia photometric band. For each relation, a second degree polynomial has been fit iteratively by clipping $1.95\sigma$ of the sources from the fit for 50 iterations until converging on the solution shown (red dashed line).}
	\label{fig:photo_transform}
\end{figure*}

Additionally, to complement the above observations, we use the expected Gaia and 2MASS magnitudes, and colors for main-sequence stars of different spectral types from \citet{pecaut2013} to predict the magnitude and color of a true companion. This will illustrate how these photometric measurements can help determine if a pair of stars are true binaries, as compared to chance alignments or spurious detections. \citet{pecaut2013} derived an empirical spectral type-color sequence for main-sequence stars in previously popular photometric bandpasses such as 2MASS, Johnson-Cousins and WISE; this empirical relationship has been updated recently to include newer bandpasses, such as Gaia DR2\footnote{\url{https://www.pas.rochester.edu/~emamajek/EEM_dwarf_UBVIJHK_colors_Teff.txt}}. As the most recent Gaia photometry provided is for DR2, and the Gaia DR2 photometric bandpasses differs from the Gaia eDR3 photometry \citep{gaiaedr3_photo}, we transform their photometry using a cross-match between Gaia DR2 and eDR3 for stars within 200 pc. As the Gaia archive does not always provide a best match between these two catalogs, we use our method from \citet{medan2021} to create a catalog of high probability matches. Using this catalog of matches, we examine relations between Gaia eDR3 and DR2 photometry for each Gaia photometric band, as shown in Figure \ref{fig:photo_transform}. For each relation, a second degree polynomial is fit iteratively by clipping $1.95\sigma$ of the sources from the fit for 50 iterations until converging on the solution shown as the red dashed line in each panel. For each band, the resulting relation between Gaia eDR3 and DR2 photometry is:
\begin{equation}\label{eq:photo_G}
	G_{eDR3} - G_{DR2} = 0.002398 x^2 - 0.02373 x + 0.02028
\end{equation}
\begin{equation}\label{eq:photo_BP}
	BP_{eDR3} - G_{DR2} = 0.07903 x^2 + 0.3368 x - 0.005311
\end{equation}
\begin{equation}\label{eq:photo_RP}
	RP_{eDR3} - G_{DR2} = 0.06902 x^2 - 0.6246 x - 0.03823
\end{equation}
where $x=BP_{DR2} - RP_{DR2}$. For the remainder of this study, the photometry from \citet{pecaut2013} is transformed in the eDR3 system using the eqs. \ref{eq:photo_G}$-$\ref{eq:photo_RP}.

\begin{figure*}[!t]
	\centering
	\includegraphics[width=1\textwidth]{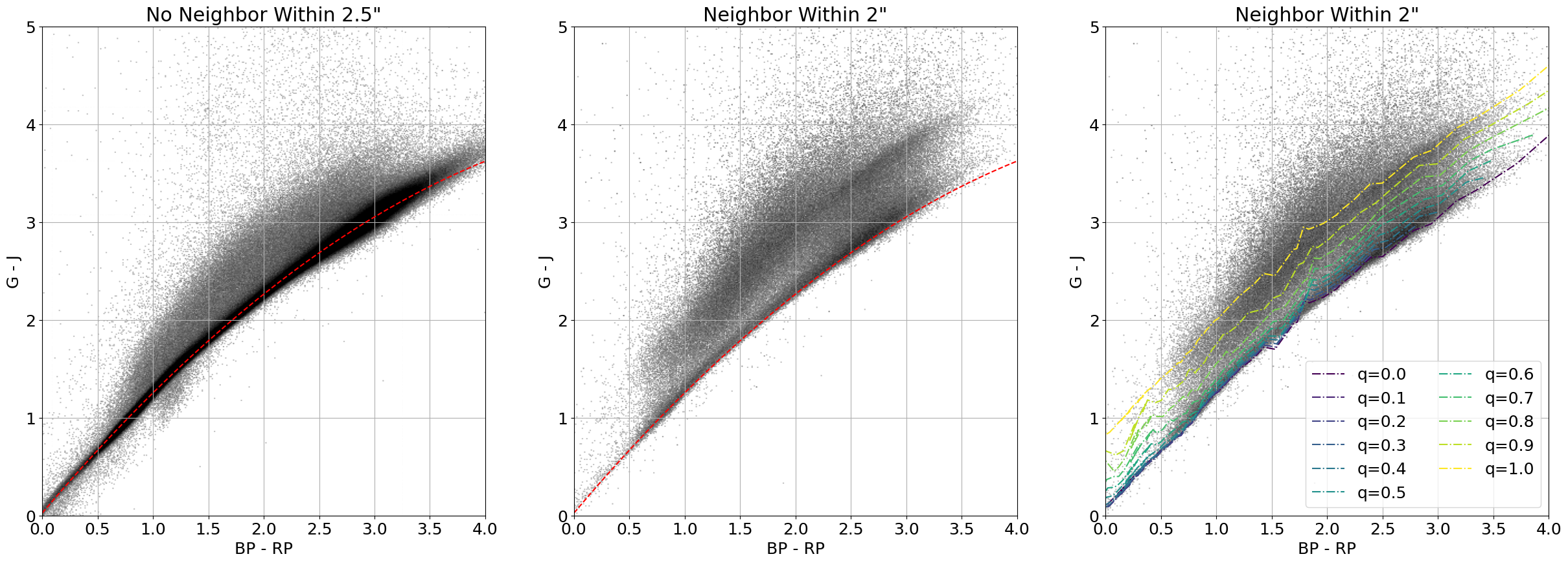}
	\caption{Color-color diagrams of $G-J$ vs. $BP-RP$ for the 200 pc Gaia eDR3 sample cross-matched with 2MASS. The left panel shows all stars that have no Gaia eDR3 neighbor within 2.5". The middle and left panels show stars than have a close neighbor within 2" listed in Gaia eDR3. The black density map in each panel is on the same logarithmic scale. The red dashed line is a 2nd degree polynomial fit to the stars that have no Gaia eDR3 neighbor within 2.5", where the fit was completed over 20 sigma-clipped iterations. The colored dashed lines overlaid on the right panel show the expected color-color relation for binary systems with various mass rations, $q$, where colors were estimated using expected colors per spectral class from \citet{pecaut2013}. When estimating expected colors, it is assumed that the Gaia $G$ photometry is only from the resolved, more massive star while the Gaia $BP$ and $RP$, and 2MASS photometry is blended from both sources.}
	\label{fig:sample_CCD}
\end{figure*}

Figure \ref{fig:sample_CCD} shows color-color diagrams of the various samples described above, where stars with no Gaia eDR3 neighbor within 2.5" are shown in the left panel and stars with a neighbor within 2" are shown in the middle panel. When comparing these two samples, it is clear that the stars with no neighbor mainly form one distinct locus which follows the expected relation between the Gaia and 2MASS bandpasses, as shown in Appendix C of \citet{gaiaedr3_photo}. To quantify this main locus, we fit a 2nd degree polynomial to this sample, with the best fit found after 20 iterations where the sample was sigma-clipped after each iteration at the $1.95\sigma$ level. This yields the following color-color relationship:
\begin{equation}\label{eq:color_rel}
	G-J = -0.1098 (BP-RP)^2 + 1.338 (BP-RP) + 0.02515
\end{equation}
shown as a red-dashed line in Figure \ref{fig:sample_CCD}.

In contrast, for the stars with a low-separation neighbor we observe two distinct loci in the color-color diagram: one that follows the expected relation between the Gaia and 2MASS colors, and a secondary locus that is shifted to the red in $G-J$ color. One explanation for this secondary locus is that it represents visual binaries in the sample that are resolved by Gaia but unresolved by 2MASS. This can be shown by modeling the expected color-color relationship for binaries with various mass ratios from the \citet{pecaut2013} calibration set, where it is assumed that the Gaia $G$ photometry is only from the more massive (resolved) member of the system, while the Gaia $BP$ and $RP$, and 2MASS photometry is the blended photometry of the (unresolved) pair. We assume that Gaia only resolves the pair in the $G$ band, because the G band is PSF fitted photometry, while the $BP$ and $RP$ bands are aperture photometry where the aperture size is larger than the area probed in this study. The predicted color-color relationships for these resolved/unresolved pairs are overlaid in Figure \ref{fig:sample_CCD}, for pairs of different mass ratios. We find that the color-color relationship for an equal mass resolved/unresolved binary is a very good fit to the location of the second locus observed in the subset of stars with a low-separation neighbor. There are additional sources that have redder $G-J$ colors than predicted by the resolved/unresolved pairs model; these could be explained by additional color-shifting effects such as reddening due to interstellar extinction or photometric reduction errors.

\section{Results}\label{sec:results4}

\subsection{Expected Relation for Visual Binaries}

\begin{figure*}[!t]
	\centering
	\includegraphics[width=0.9\textwidth]{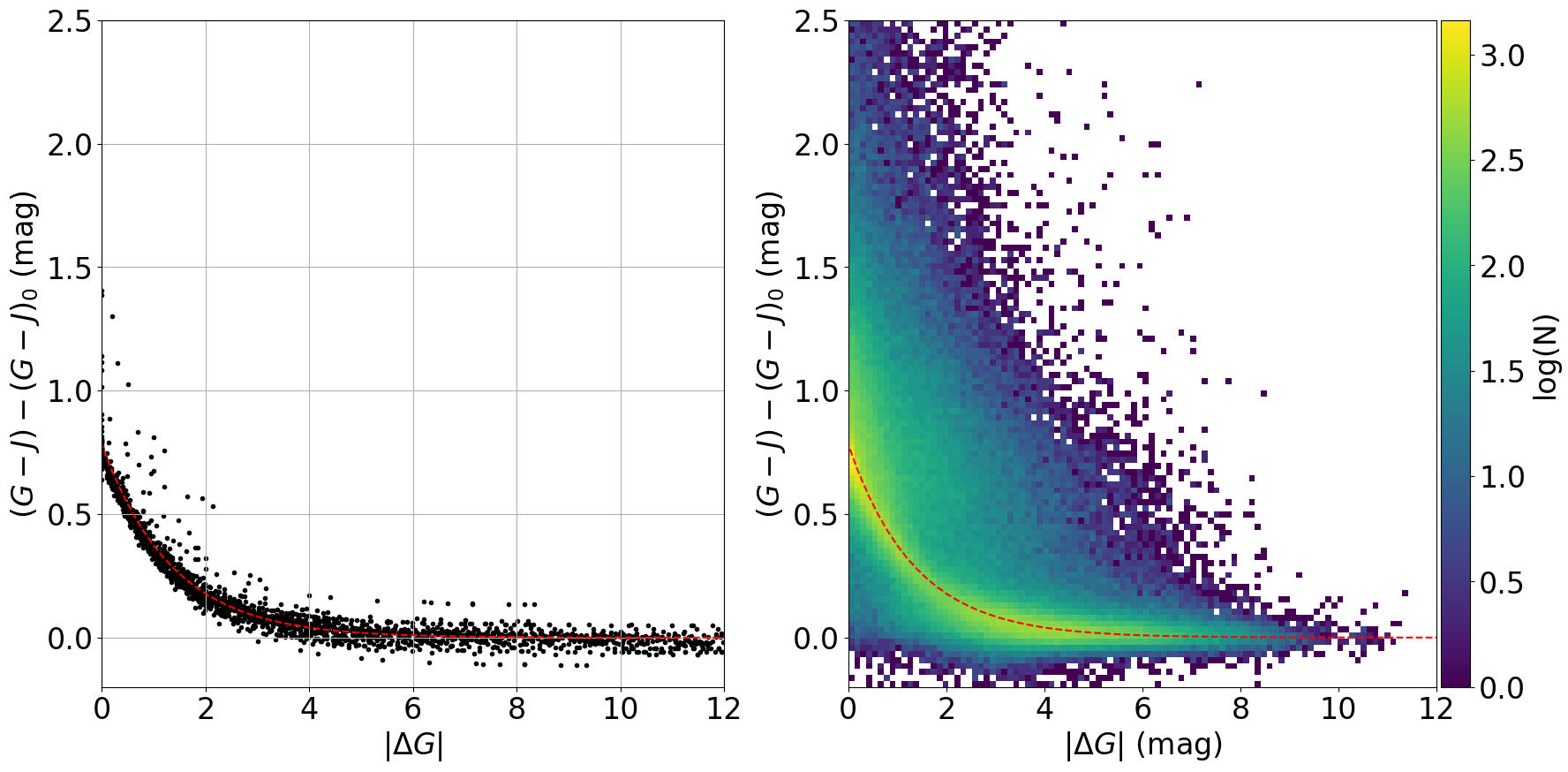}
	\caption{The expected (left panel) and observed (right panel) $G-J$ excess as a function of $|\Delta G|$ between a Gaia eDR3 star and its low-separation neighbor. For the expected excess, the expected photometry from \citet{pecaut2013} is used to create the black data points. Here the excess is defined as the difference between the $G-J$ color with the 2MASS photometry artificially blended and the expected $G-J$ color from eq. \ref{eq:color_rel} based on the artificially blended $BP-RP$ color for the sources. For the observed excess, we define the excess as the difference between the observed $G-J$ color and the expected $G-J$ color based on the star's $BP-RP$ color (eq. \ref{eq:color_rel}; $(G-J)_0$). The dashed red line is an exponential function fit to the expected $G-J$ excess data, where the expected relation matches the observed relation well.}
	\label{fig:expected color_excess}
\end{figure*}

Even though it is expected that there will be a color excess if a pair of stars are at a separation less than the PSF size of 2MASS and are visual binaries in Gaia, it is also true that there will be a similar excess if the neighbor is a random field star. Depending on the spectral types of the components however, there should be a predictable relationship for the $G-J$ excess of a visual binary as a function of the magnitude difference, $|\Delta G|$, between the components. This is demonstrated in the left panel of Figure \ref{fig:expected color_excess}, where the expected photometry from \citet{pecaut2013} is used to create a relationship between the observed color excess $\Delta(G-J)_0 =  (G-J) - (G-J)_0$ and the magnitude difference in Gaia G band of the (resolved) components, $|\Delta G|$. Here, the $G-J$ color excess is defined as the difference between the $G-J$ color of the blended (unresolved) pair, and the predicted color $(G-J)_0$ for a single star with the same $BP-RP$ color as the blended pair, based on eq. \ref{eq:color_rel}. The way this  color excess relationship is calibrated assumes that the stars are both main sequence (MS) stars and that they are at the same distance. The black data points in the left panel of Figure \ref{fig:expected color_excess} are these excesses calculated for each spectral type listed in the \citet{pecaut2013}, and then for every possible pair of stars from these spectral types. This then should probe the possible values for all main-sequence spectral types and mass ratios, but is not necessarily representative of the underlying magnitude and mass ratio distribution in the 200 pc sample, so this expected relation will only be used for visualization purposes.

The right panel of Figure \ref{fig:expected color_excess} shows the color excess distribution for the close pairs in our Gaia eDR3 200 pc sample. For this observed excess plot, we define the color excess as the difference between the observed $G-J$ color and the predicted color, $(G-J)_0$, for a single star of the same $BP-RP$ color (eq. \ref{eq:color_rel}). We find that the distribution from the model (left panel Figure \ref{fig:expected color_excess}) can be fit with the function of the form:
\begin{equation}
	\Delta(G-J)_0 = 0.7799 e^{-0.7382 |\Delta G|}
\end{equation}
This relationship is shown as the dashed red line in Figure \ref{fig:expected color_excess}. One can see that the distribution for the observed eDR3 sources follows the same trend as the one expected from the calibration subset.

We do recognize that this relationship is not entirely reflective of the candidate sample, as we also expect there to be white dwarfs (WDs) and, to a lesser extent, giants within 200 pc. Because of this, we do expect that within our sample there are going to be binaries that are not MS-MS pairs, but pairs with a WD or giant component, or instances where both components are WDs or giants. While we do not expect these types of systems to make up the majority of our sample, we can examine how the color excess relationship behaves for MS-WD binaries, as these are the most common type of white dwarf binary \citep{holberg2016}. 

\begin{figure}[!t]
	\centering
	\includegraphics[width=0.5\textwidth]{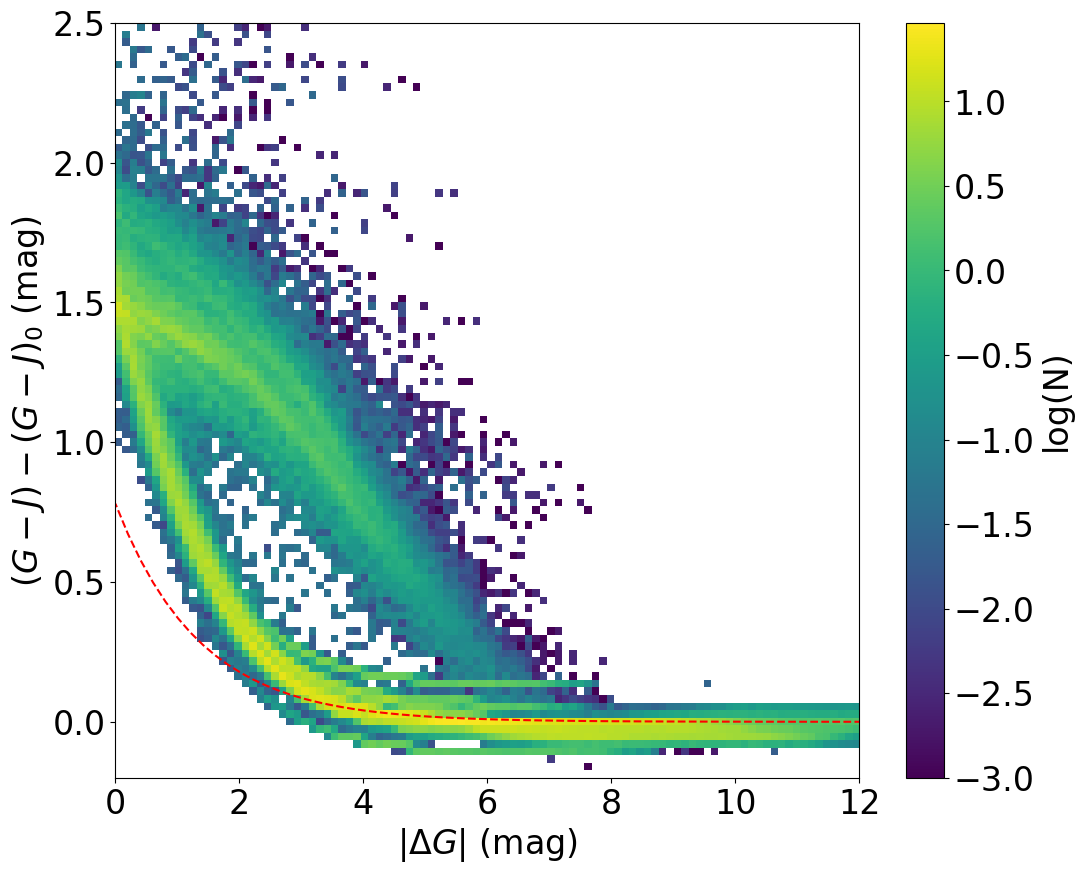}
	\caption{The expected $G-J$ excess for WD-MS binaries. Here white dwarf photometry from \citet{fusillo2021} is combined with random samples of photometry from \citet{pecaut2013} and bootstrapped over 100 iterations to get the average distribution shown. The dashed red line is an exponential function fit to the expected excess from MS$+$MS binaries, as shown in the left panel of Figure \ref{fig:expected color_excess}.}
	\label{fig:WD_expect}
\end{figure}

To accomplish this, we us the Gaia eDR3 white dwarf catalog from \citet{fusillo2021}, where we select the white dwarfs from their catalog within our 200 pc sample. As stated in \citet{fusillo2021}, only 3\% of the white dwarf catalog is expected to be contaminated with MS-WD binaries or cataclysmic variables, so we assume the catalog consists primarily of white dwarfs (either single or double degenerate systems) and we will need to artificially blend them with main sequence stars for our purposes. We do this by again using the photometry from \citet{pecaut2013}. Here we randomly select photometry from \citet{pecaut2013} such that the sample is the same size as the selected white dwarf sample and such that the selected $M_G$ from the \citet{pecaut2013} selection has the same underlying distribution as the overall 200 pc sample. We then blend the $BP$, $RP$, and $J$ magnitudes, and calculate the $G-J$ as above. This process is repeated 1000 times to bootstrap the average, expected distribution. This distribution is shown in Figure \ref{fig:WD_expect}. Here we see that the distribution is significantly different for small $|\Delta G|$ values, as compared to the relationship expect for MS-MS binaries. These larger $G-J$ excesses could describe some of the excess of stars in the observed distribution with $ (G - J) -  (G - J)_0 > 1$.

\subsection{Expected Relation for Field Stars}\label{sec:models}

\begin{figure*}[!t]
	\centering
	\includegraphics[width=\textwidth]{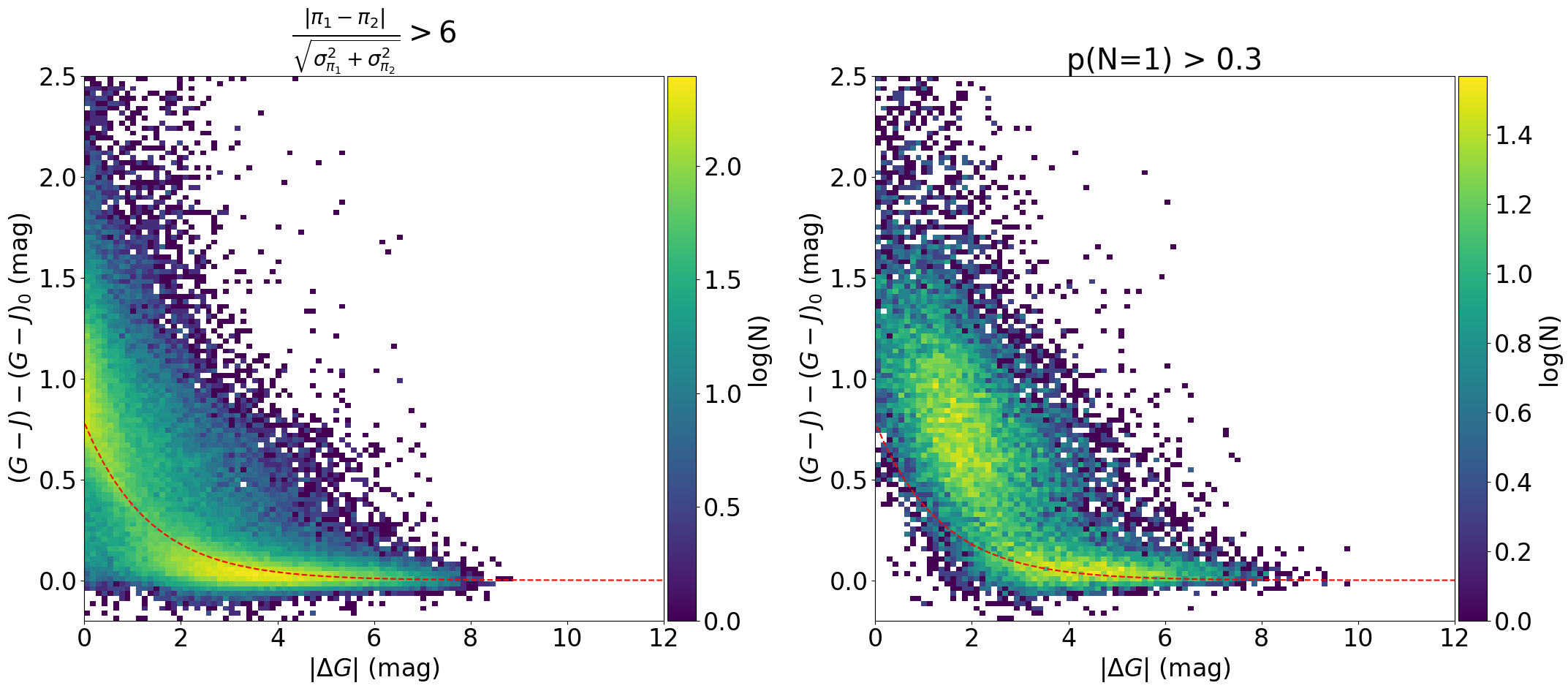}
	\caption{The observed $G-J$ excess as a function of $|\Delta G|$ between a Gaia eDR3 star and its low-separation neighbor for stars with significant parallax differences (left panel) and high Poisson probability (right panel). The excess is defined as the difference between the observed $G-J$ color and the expected $G-J$ color based on the star's $BP-RP$ color (eq. \ref{eq:color_rel}; $(G-J)_0$). The dashed red line is an exponential function fit to the expected $G-J$ excess data.}
	\label{fig:expected color_excess_field}
\end{figure*}

While we can directly model the expected relationship of $G-J$ excess vs. $|\Delta G|$ for unresolved binary stars using the expected relative photometry of main sequence stars and white dwarfs, it is not as straightforward to do this for unrelated field stars that happen to be chance alignment neighbors to a Gaia eDR3 200 pc star. We can examine, observationaly, what the expected $G-J$ excess distribution of fields stars to look like by using the sample of likely chance alignments determined from significant parallax differences and high Poisson probabilities (Figure \ref{fig:expected color_excess_field}). This also has the benefit of giving the distribution of chance alignments that has the same magnitude and sky distribution we would expect from the specific 200 pc sample, something that our models for visual binaries above cannot easily accomplish. In the distributions we observe the following. 

First, for both samples of field stars it is clear that we don't expect as large of a portion of the pairs to have small $|\Delta G|$ values. \textcolor{black}{For example, when comparing the right panel of Figure \ref{fig:expected color_excess} (observed distribution) to the left panel of Figure \ref{fig:expected color_excess_field} (pairs with significant parallax differences), while both distributions have a large portion of stars with $|\Delta G| < 1$ and $0.5 < (G-J) - (G-J)_0< 1$, the observed distribution has a higher proportion of the overall sample in this region than the presumed \textcolor{black}{chance} alignments. Despite this difference in proportion between the two samples in this regime ($|\Delta G| < 1$ and $0.5 < (G-J) - (G-J)_0< 1$),} we do see that the \textcolor{black}{overall shape of the} $G-J$ excess vs. $|\Delta G|$ \textcolor{black}{distribution} for \textcolor{black}{these} chance alignments when there is a secondary parallax, is not significantly different than what we expect for true MS-MS binaries. Also in this \textcolor{black}{$|\Delta G| < 1$} regime, we do see that a non-negligible number of the chance alignments have $G-J$ excess values $>1$. It should be noted that in this regime we \textit{do} see some excess though, indicating that these detections are not bogus. \textcolor{black}{Such large excesses then could be distant background stars with high reddening values, which means the $G-J$ excess will be even greater for some observed magnitude difference in the optical, and could account for the excess of high $G-J$ excess stars seen in the overall observed distribution right panel of Figure \ref{fig:expected color_excess}. Finally, for the pairs with larger magnitude differences ($2< |\Delta G| < 4$), we note a slight difference between the observed distribution and the distribution of pairs with significant parallax differences. For the latter, we find the $G-J$ excess is nearer to $0$ than to the expected distribution for MS-MS binaries (red dashed line), while the observed distribution is centered more on this expected relationship. This slight difference could be helpful in differentiating true binaries and chance alignments in this regime.}

\textcolor{black}{These comparisons are} not the same for the high Poisson probability field stars (where nearly half don't have parallax measurements), where for $|\Delta G| < 2$ the distribution is significantly different than what we expect for MS-MS true binaries. This is most likely due to these stars being relatively distant background stars with high reddening values, \textcolor{black}{similar to what is seen in the chance alignments with large parallax differences}. This does match the larger excesses observed in the right panel in Figure \ref{fig:expected color_excess}, indicating that within our candidate sample there is contamination from chance alignments with background sources. We also find though that these chance alignments without parallax measurements have similar $G-J$ excesses as the MS-WD binaries (Figure \ref{fig:WD_expect}). Overall, this demonstrates that we may not be able to differentiate field stars and true binaries in this parameter space alone.

\subsection{Determining Likelihood Distributions}\label{sec:prob_dists}

\begin{figure*}[!t]
	\centering
	\includegraphics[width=\textwidth]{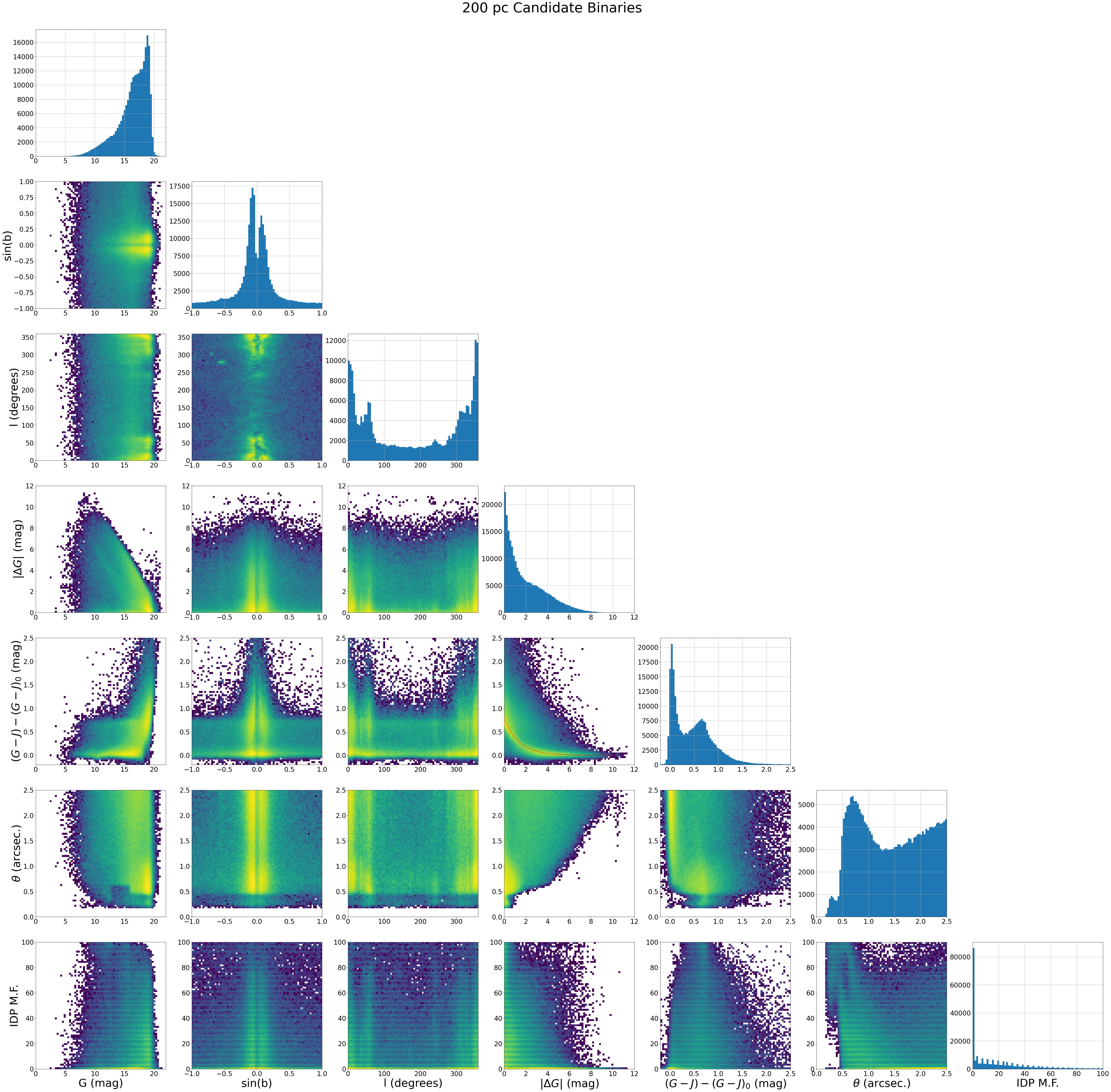}
	\caption{Corner plot for all 200 pc stars with a close Gaia neighbor show the distributions in $G$ magnitude of the primary ($G$), sine of the Galactic latitude ($sin(b)$), Galactic longitude ($l$), $|\Delta G|$, $G-J$ excess ($(G - J) -  (G - J)_0$), \textcolor{black}{angular separation ($\theta$) and the \texttt{ipd\_frac\_multi\_peak} value from \textit{Gaia} (IDP M.F.)}. All 2D distributions are log-scaled to better show low signal details in the distributions. All 1D distributions show the number linearly scaled.}
	\label{fig:corner_plot_cands}
\end{figure*}

\begin{figure*}[!t]
	\centering
	\includegraphics[width=\textwidth]{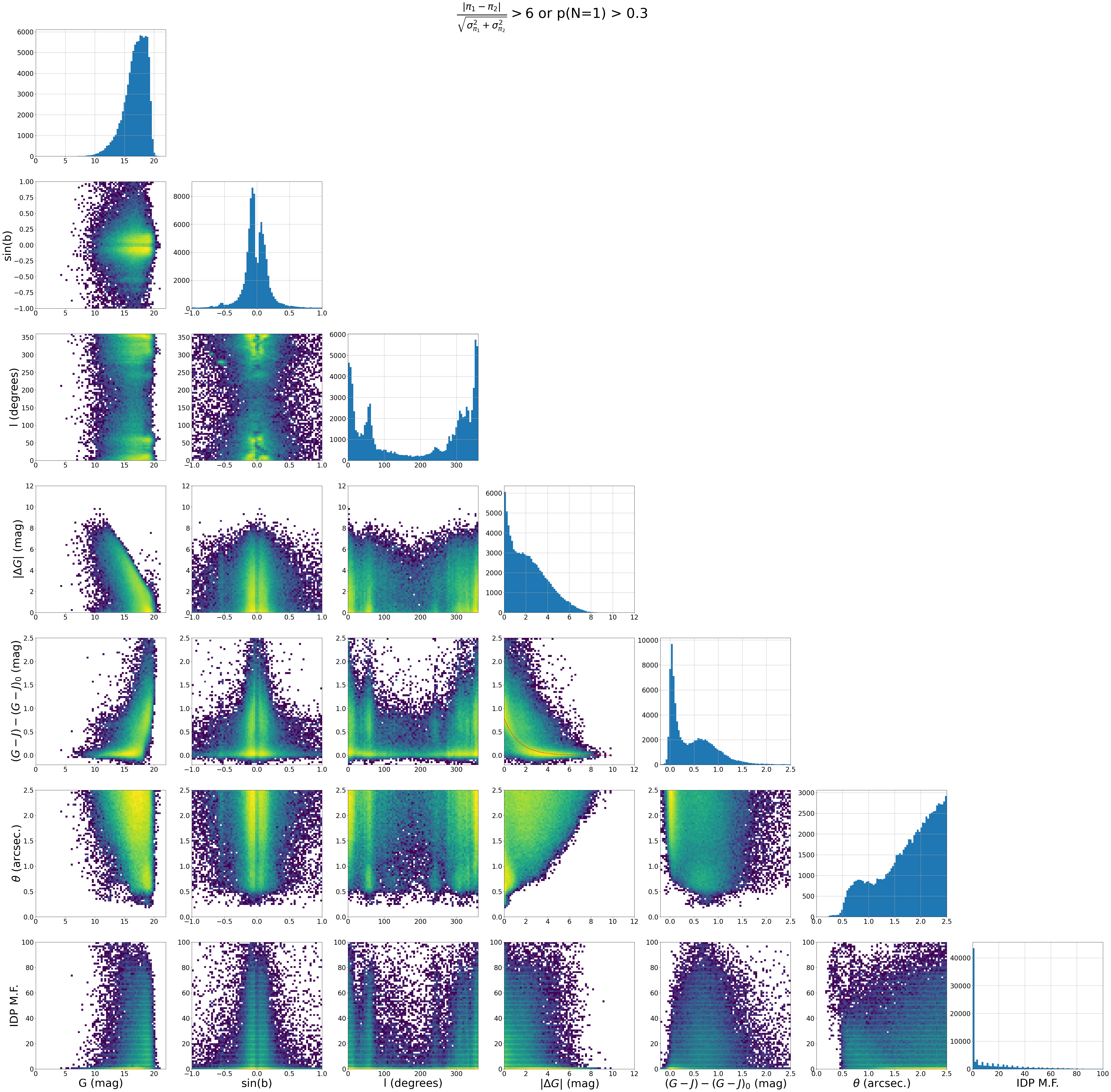}
	\caption{Same as Figure \ref{fig:corner_plot_cands} except only for pairs that have significant parallax differences or high Poisson probability of chance alignment, which mean that the "companion" is a background source.}
	\label{fig:corner_plot_field}
\end{figure*}

As demonstrated in the last two sections, it appears difficult to determine if a pair of stars are true binaries based on $G-J$ excess alone. Indeed this likely depends additionally on the magnitude and location on the sky of the pair of stars, \textcolor{black}{as well as parameters related to the separation of the two stars}. Specifically, we examine the difference in the distribution of candidates and likely field stars when considering the $G$ magnitude of the primary, sine of the Galactic latitude, Galactic longitude, $|\Delta G|$, $G-J$ excess\textcolor{black}{, angular separation between the stars and the \texttt{ipd\_frac\_multi\_peak} value from \textit{Gaia}}. The corner plots for the 200 pc candidate binaries and the likely chance alignments are shown in Figures \ref{fig:corner_plot_cands} and \ref{fig:corner_plot_field}, respectively. Here we see that there are more stark differences between the two samples. 

In this higher dimensional space it is clear that we expect most chance alignments with $|\Delta G| < 2$ and $(G - J) -  (G - J)_0 \approx 0.75$ to be faint ($G > 17$) and be at low Galactic latitude. For the distribution of all 200 pc stars with close Gaia companions, we see there are many pairs that are brighter and at higher Galactic latitudes, indicating that in this regime we are more likely to differentiate true binaries from field stars based on their $G-J$ excess. Additionally, it seems there is a difference between the likely field stars and candidates at larger values of $|\Delta G|$, which also corresponds to true binaries of lower mass ratios. For $0 < (G - J) -  (G - J)_0  < 0.5$ and $|\Delta G| > 2$, we see that in the chance alignment distribution that, again, these pairs are more likely to be lower Galactic latitudes and relatively faint. Especially for the latter, it seems that even when the primary is brighter, the secondary is consistently near the faint limit of Gaia, i.e. when the primary has a $G \approx 15$, it is likely that $| \Delta G| \approx 5$. While these trends are also seen in the full set of close pairs (likely representing the part of the distribution that contains chance alignments), there is also an excess of close pairs at higher Galactic latitude (for this range of $G-J$ excess and $|\Delta G|$) and it is clear you can have pairs with larger $|\Delta G|$ values where the secondary doesn't fall near the faint limit; this excess most likely represents the true binaries.

\textcolor{black}{Also, we find that the parameters related to the separation of the two stars, angular separation and \texttt{ipd\_frac\_multi\_peak}, show stark differences between the candidate binaries and chance alignments. For the distribution of angular separations for the chance alignments, the distribution is dominated by a rising linear trend that is expected for chance alignments \citep{hartman2020,elbadry_2021}. While this is still present in the candidate binaries, there is a much larger peak at shorter separations ($ 0.5 < \theta < 1$") than in the chance alignment distribution. We do note that this peak at $ 0.5 < \theta < 1$" is present in the chance alignment distribution but (1) it is at not as prominent as the linear trend and (2) seems to be concentrated at the faint end of the magnitude distribution. Overall, we attribute this additional peak in the chance alignment angular separation distribution at $ 0.5 < \theta < 1$" to spurious solutions. Spurious solutions in \textit{Gaia} can occur for close source pairs depending on the scan angle for the observation and will result in meaningless parallax and proper motion values \citep{gaiaedr3_catalog_valid}. The number of spurious detections is expected to increase at fainter magnitudes and for certain regions on the sky mostly near the Galactic center \citep{gaiaedr3_catalog_valid}, which match well with the trends seen in the chance alignment sample. Finally, the pairs that are within this peak seem to only be apart of the sample with significant parallax differences. If these are spurious detections this criteria may be invalid. Because these sources are concentrated in the Galactic center, where later in this paper we will remove most binaries due to high contamination rates, the addition of these sources with spurious solutions will not effect the results of this work.}

\textcolor{black}{Finally, the use of the \texttt{ipd\_frac\_multi\_peak} parameter seems to differentiate between candidate binaries and chance alignments in some areas of the parameter space. While the value is high for both candidate binaries and chance alignments in the Galactic plane where the stellar density is highest, it is very unlikely to see a high \texttt{ipd\_frac\_multi\_peak} value ($>20$) at higher Galactic latitudes. It is in this regime this parameter may be most useful to detect genuine binaries. We do note that at larger angular separations ($\theta > 1$"), we expect that even true binaries will have small or near zero \texttt{ipd\_frac\_multi\_peak} values \citep{Tokovinin2023}, so this parameter will most likely not be useful in this regime.}

As this higher dimensional space seems to better differentiate true binaries from chance alignments, we follow a procedure similar to the one in \citet{elbadry_2021} to get \textcolor{black}{a value analogous to a} likelihood of a pair of stars being a chance alignment\textcolor{black}{, which in this paper we will refer to as a ``contamination factor"}. To estimate probability densities from the parameter space described above, we use a Gaussian kernel density estimate (KDE), as implemented in \texttt{scikit-learn} \citep{scikit-learn}. We normalize the data to the bounds of the plots on Figures \ref{fig:corner_plot_cands} and \ref{fig:corner_plot_field} and use a kernel bandwidth of $0.02$. \textcolor{black}{Here, the data is normalized to bounds of the plots so all parameters have a similar dynamic range near 1. This is done as the Gaussian kernel size is constant in all dimensions, so once normalized the smoothing applied will be, relatively, similar in all dimensions.} We then use the densities from the two distributions, $N_{cand}(\overrightarrow{x})$ for all 200 pc candidates (Figure \ref{fig:corner_plot_cands}) and $N_{chance}(\overrightarrow{x})$ for likely chance alignments (Figure \ref{fig:corner_plot_field}), to get the \textcolor{black}{contamination factor that measures how likely a candidate binary is to be a chance alignment:}
\begin{equation}\label{eq:likelihood4}
	L = \frac{N_{chance}(\overrightarrow{x})}{N_{cand}(\overrightarrow{x})}
\end{equation}
\textcolor{black}{In the above, $N(\overrightarrow{x})$ is meant to represent the 7 dimensional distributions smoothed with the KDE, where the 7 parameters for $\overrightarrow{x}$ are the ones shown in the corner plots in Figures \ref{fig:corner_plot_cands} and  \ref{fig:corner_plot_field}. Also, it should be noted that the above contamination factor, $L$,} is not strictly a probability, as not all values fall between 0 and 1, but once calibrated the above should work as a parameter to determine if a population of candidates are true binaries to some detection limit and with some false positive rate.

\subsection{\textcolor{black}{Determining} Likely Binaries}

To determine the most likely binaries, we need to calibrate the \textcolor{black}{contamination factor}, $L$, described above. \textcolor{black}{This is needed as $L$ is not strictly a probability, so selecting some value will not mean that we are selecting binaries with some numerical value of confidence. Instead, we must determine some way to quantify the level of contamination based on some cut in $L$ to the sample.}

\textcolor{black}{One way to do this is to \textcolor{black}{assume} that the binaries should have some distribution across the sky. If we assume that the likely binaries should have the same distribution as the 200 pc stars, than any regions on the sky that have excesses relative to this expected distribution signify some level of contamination from chance alignments. Additionally, any deficit compared to the expected distribution signifies that our catalog is not complete in this region on the sky. So, to model the expected background we do the following.}

\begin{figure*}[!t]
	\centering
	\gridline{\fig{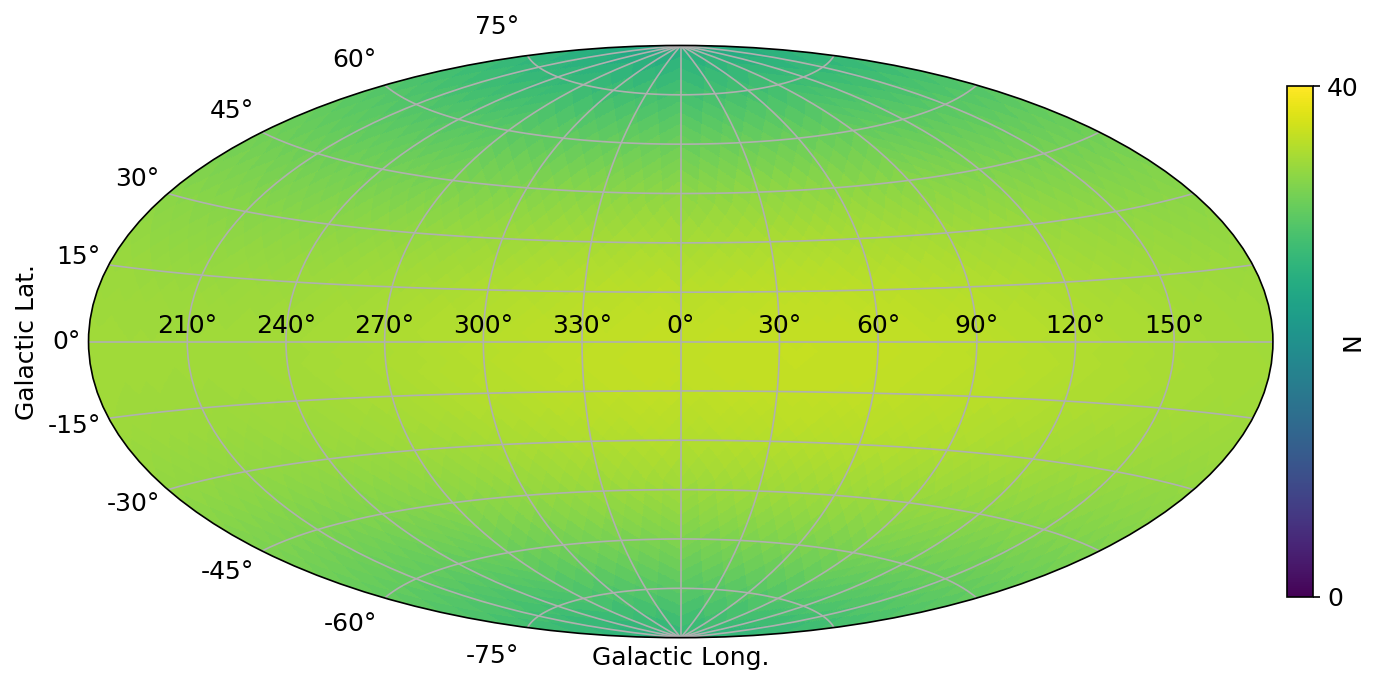}{0.45\textwidth}{(a) Expected Distribution} \fig{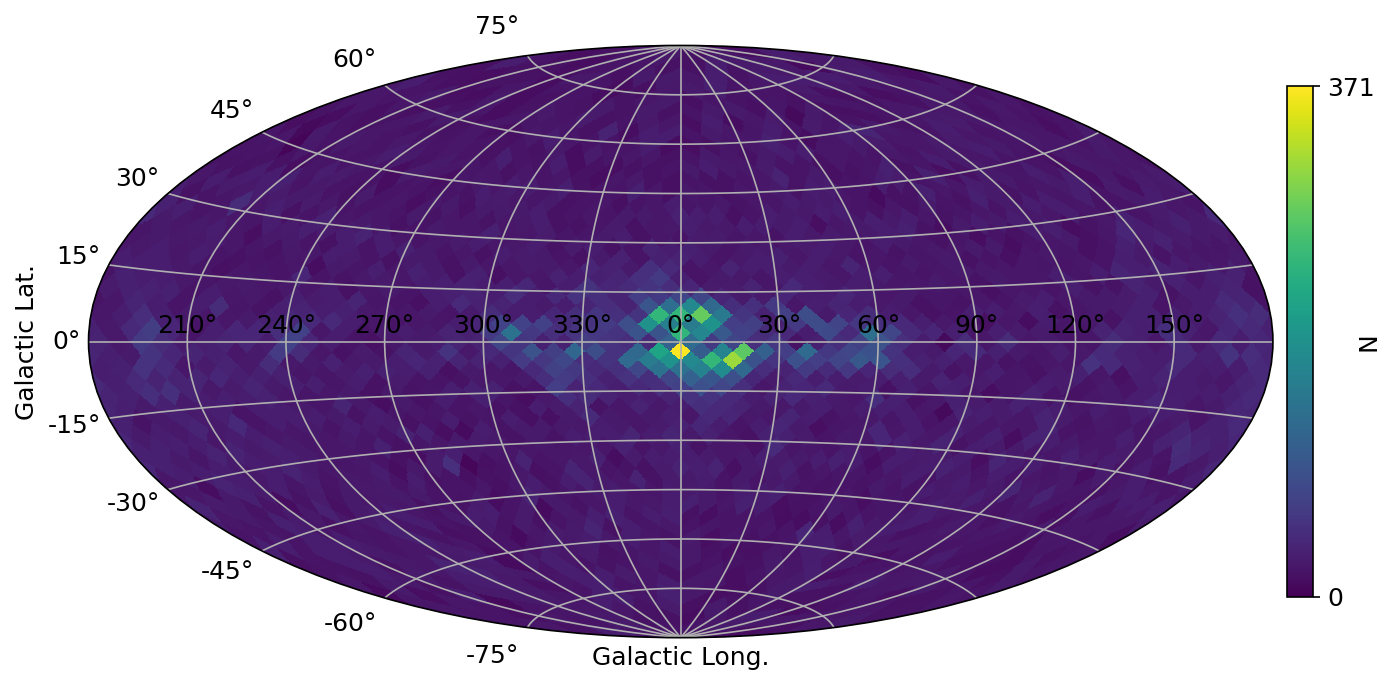}{0.45\textwidth}{(b) Likely Binaries}}
	\gridline{\fig{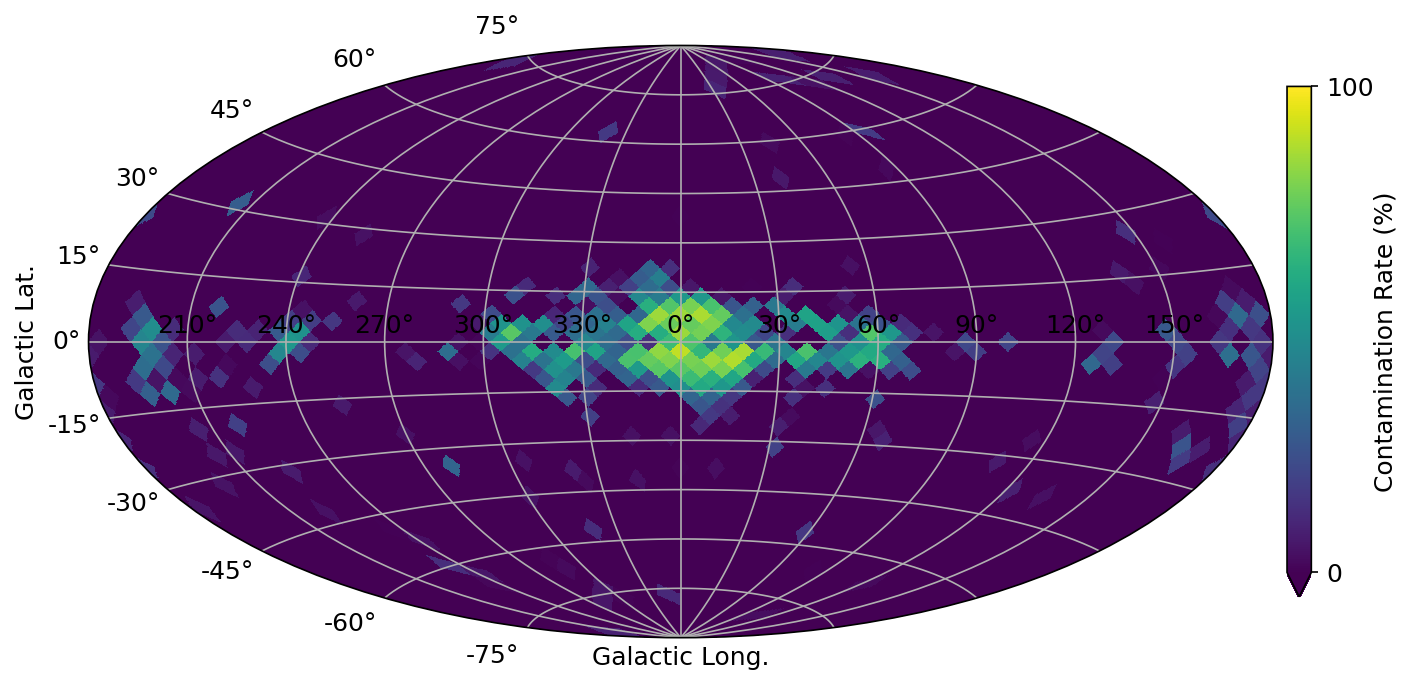}{0.45\textwidth}{(c) Contamination Rate} \fig{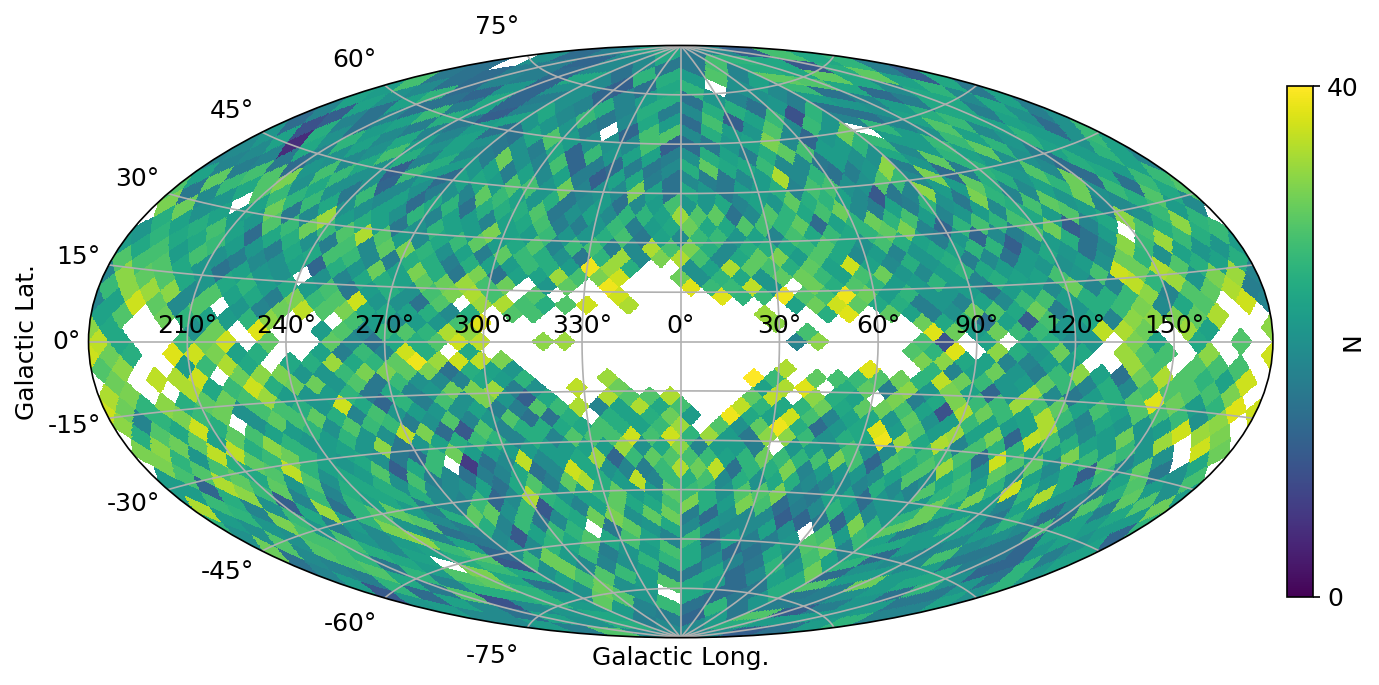}{0.45\textwidth}{(d) Likely Binaries w/ Contam. Rate \textcolor{black}{$<10\%$}}}
	\caption{Sky plots in Galactic coordinates for \textcolor{black}{the expected distribution of binaries from our model based on the 200 pc sample (top left),} all likely binaries (top \textcolor{black}{right}), contamination rate in likely binaries (\textcolor{black}{bottom left}) and likely binaries in regions where the contamination rate is \textcolor{black}{$<10\%$} (bottom \textcolor{black}{right}).  The full list of candidate binaries is given in Table \ref{tab:candidate_binaries}.}
	\label{fig:sky_plots}
\end{figure*}

\textcolor{black}{To model the expected distribution, we assume that the sky density of stars follows:
\begin{equation}
	\rho(l, b) =  \left[A \ sin(c \times l + \omega) + C\right] \times (a_2 b^2 + a_1 b + a_0)
\end{equation}
In the above, $l$ and $b$ are Galactic longitude and latitude, such that the expected stellar density is sinusoidal in the longitude direction and parabolic in the latitude direction. We fit the above relationship to the sky density of the entire 200 pc sample in healpix bins with $n_{side} = 14$. This fit is done iteratively, where for each iteration we remove healpix bins that are $>2\sigma$ from the mean difference between the observations and the model. This is done to ignore high density regions near the Galactic center which are a result of things like spurious solutions of some stars in the 200 pc sample. The resulting fit to this expected distribution is shown in Figure \ref{fig:sky_plots}a. As a note, this is scaled to the binary population, which will be discussed below. This distribution is roughly uniform, with some slight deficits at the Galactic poles and slight excesses towards the Galactic center.}

\textcolor{black}{With this expected background distribution, we then find the contamination rate across the sky. To do this, we look at a subset of candidate binaries where $L < L_{cut}$. The number of binaries that meet this criteria are the likely binaries. We then scale the expected sky distribution to this distribution of likely binaries based on the healpix bins with $|b| > 80^\circ$, as we assume that the binaries in the catalog should be complete in the Galactic poles due to the low stellar density in this region. With this scaled expected distribution, the contamination rate in a healpix bin is:
\begin{equation}
	C_i = \frac{N_{i, binary} - N_{i, exp}}{N_{i, binary}}
\end{equation}	
where $N_{i, binary}$ is the number of likely binaries and $N_{i,exp}$ is the expected number based on the scaled distribution. The number of contaminates for $L < L_{cut}$ is $\sum_i C_i \times  N_{i, binary}$ for $C_i >0$. We do note that there are healpix bins with $C_i <0$, which are regions on the sky \textcolor{black}{where} our binary catalog is not complete. To assess the usefulness of our catalog though, we determine a ``clean" sample at each iteration as well, which is the number of likely binaries in regions where $C_i < 0.1$.}

\begin{figure*}[!t]
	\centering
	\includegraphics[width=\textwidth]{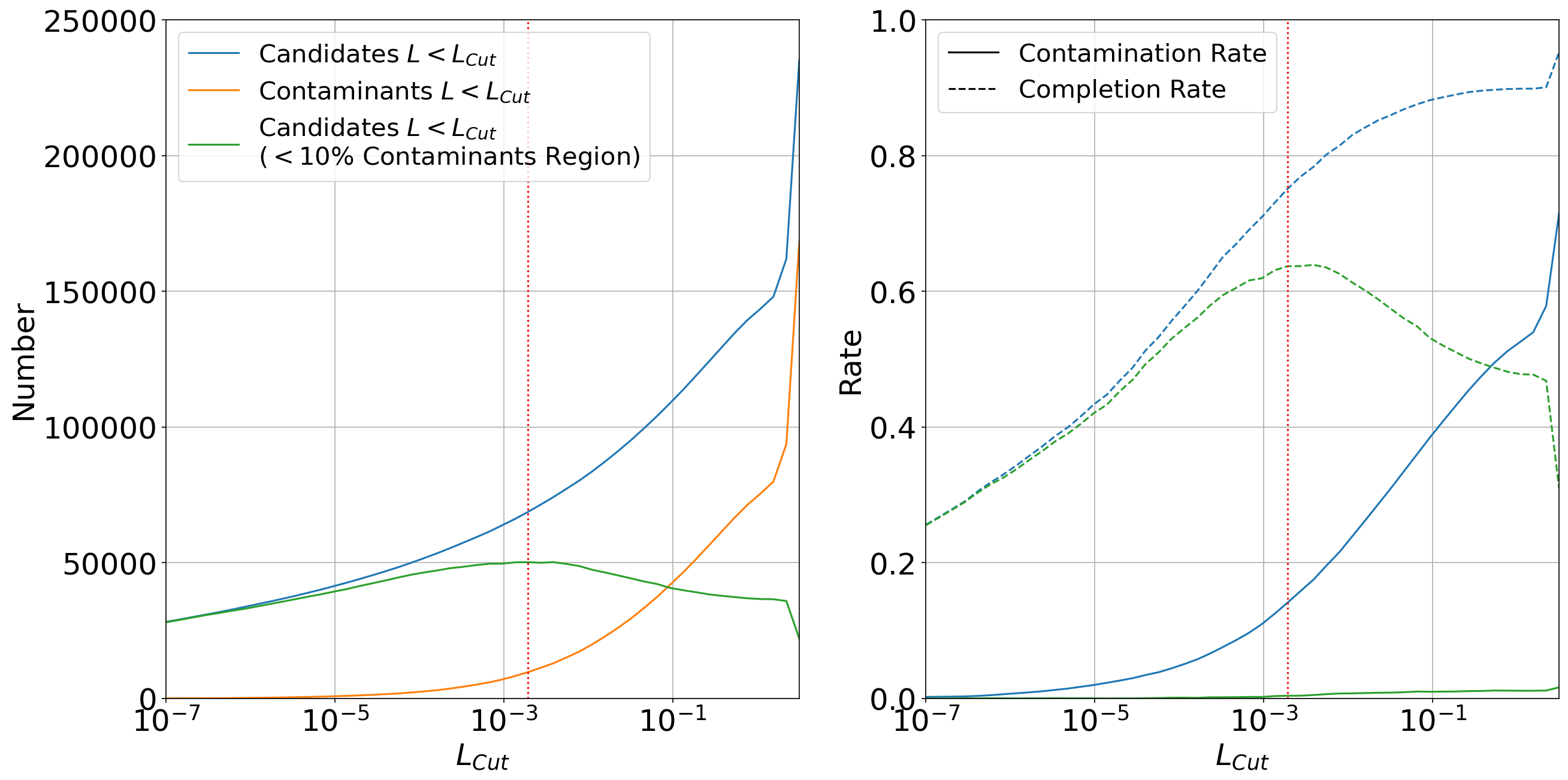}
	\caption{\textcolor{black}{\textbf{Left Panel:} The number of candidate binaries (blue line) and likely contaminants (orange line) given some cut in contamination factor, $L_{cut}$, of being a chance alignment (eq. \ref{eq:likelihood4}). Here, the candidates are all candidate binaries that have $L < L_{cut}$, while the contaminants are the number of candidates greater than what is expected for a 200 pc sample of stars based on their sky distribution. Finally, the green line shows the resulting number of candidate binaries if high contamination regions ($>10\%$) on the sky are excluded. The red dotted line shows the ideal value of $L_{cut} = 0.00193$. \textbf{Right Panel:} The contamination rate (solid lines) and completion rate (dashed lines) for all candidate binaries with $L<L_{cut}$ (blue lines) and candidate binaries with $L < L_{cut}$ in low contamination regions ($< 10\%$; green lines). The red dotted line shows the ideal value of $L_{cut} = 0.00193$.}}
	\label{fig:contam_complete}
\end{figure*}

\textcolor{black}{Following this procedure, the number of all likely binaries (blue line), contaminants (orange line) and number of likely binaries in low-contamination regions ($C_i < 0.1$; green line) as a function of $L_{cut}$ are shown in left panel of Figure \ref{fig:contam_complete}. Here we see that as we increase $L_{cut}$, we increase the number of likely binaries we are sensitive to at the cost of a increased number of contaminants. Similarly, we see an increase in the clean sample of likely binaries until a point that the high contamination regions dominate and we begin to loose usable binaries within the catalog. Another way to analyze this is by looking at the contamination and completion rates of the samples (right panel of Figure \ref{fig:contam_complete}). To calculate these rates, we assume that all regions with $C_i < 0$ have no contaminants, so only probe the completion of our sample. As expected for the full sample of likely binaries, as we increase $L_{cut}$ we get an increase in completion at the expense a higher contamination rate. By removing the binaries in high contamination regions though, we can get a \textcolor{black}{consistently} low contamination rate. But, after the turnover point this comes at a cost of a lower completion rate.}
	
\textcolor{black}{It is at this turnover point that we consider the ideal $L_{cut}$ value for our final catalog, such that it contains binaries with $L < 0.00193$ (red dotted line Figure \ref{fig:contam_complete}). This results in a catalog of 68,725 likely binaries, 50,230 of which are part of the clean sample in low contamination regions. At this cut, the resulting sky distribution for the full catalog is shown in Figure \ref{fig:sky_plots}b, the contamination rate across the sky in Figure \ref{fig:sky_plots}c and the sky distribution of the clean sample in Figure \ref{fig:sky_plots}d. Here we can see that most of the high contamination regions are in the direction of the Galactic center.}

\textcolor{black}{Based on this above analysis, we expect a contamination rate of 14.1\% in the full catalog and 0.4\% for the clean sample. Also when comparing the expected distribution from our model (Figure \ref{fig:sky_plots}a) and for the clean sample (Figure \ref{fig:sky_plots}d), we see that for many regions on the sky our catalog has fewer binaries than expected indicating the catalog is not complete. If we assume that all regions with $C_i < 0$ have no contaminants, we can determine the completeness of our sample. From this, we expect that our full catalog is 75\% complete and our clean sample is 64\% complete.}

\begin{figure*}[!t]
	\centering
	\includegraphics[width=\textwidth]{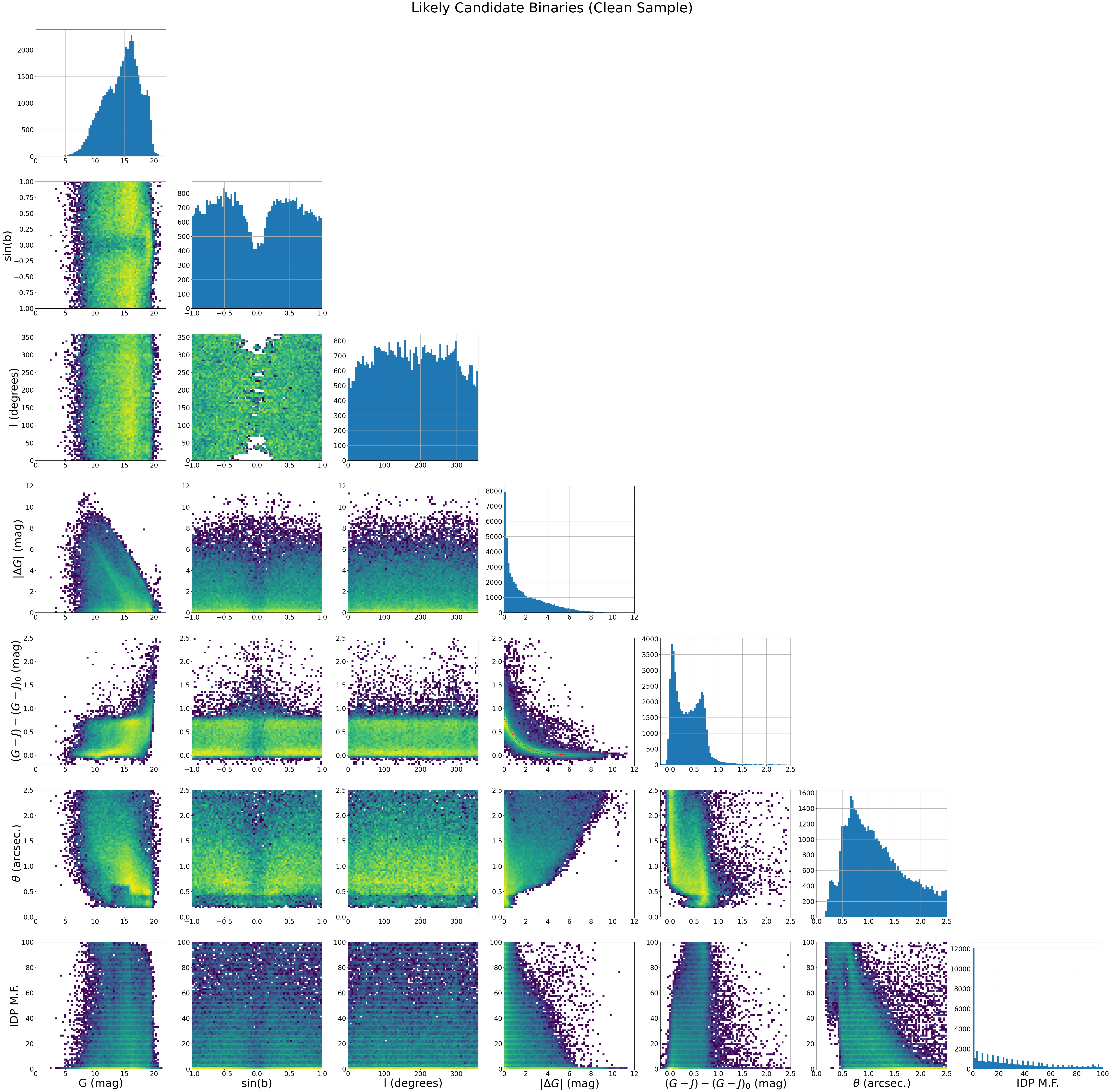}
	\caption{Same as Figure \ref{fig:corner_plot_cands} except for the clean sample of likely binaries.}
	\label{fig:corner_plot_likely}
\end{figure*}

The resulting corner plot for the clean subset of likely binaries is shown in Figure \ref{fig:corner_plot_likely}. The distribution follows the expected $G-J$ excess vs. $|\Delta G|$ relationship for MS-MS binaries very well, though there remains a small number of \textcolor{black}{likely} binaries with $G-J$ excesses much greater than what is expected from this relationship. As shown previously, these candidates could still be true binaries if they happen to be MS+WD systems (see Figure \ref{fig:WD_expect}). Overall, it is possible that we are probing a fairly large range of $|\Delta G|$ because we can better differentiate chance alignments from binaries (both MS+MS and MS+WD) for bright stars and for stars at higher Galactic latitude. Here, higher $G-J$ excess values are less common for chance alignments, but also the small color excesses for large values of $|\Delta G|$ are less common. \textcolor{black}{Also, when examining the angular separation distribution for the clean sample, we see that the linear trend associated with chance alignments in not present. This is reassuring and helps support that the overall contamination rate of the clean sample is indeed low.}

We also note \textcolor{black}{a peak in} faint primaries \textcolor{black}{near the \textit{Gaia} magnitude limit} at lower Galactic latitudes. \textcolor{black}{These are most likely regions where the contamination rate is near 10\%, which are concentrated near the Galactic plane and Galactic center, such that they make it into the clean sample but still may have a larger contamination rate that most other positions on the sky. For users of this catalog, such issues should be kept in mind and additional cleaning of the catalog may be warranted depending on the application of the catalog.}

With the above considered, the list containing all candidate binaries is found in Table \ref{tab:candidate_binaries}, where columns in this table with the subscript 1 identifies the primary star, which is the star used to measure the $G-J$ excess in the probability distributions, and the subscript 2 identifies the neighboring secondary star. In this table we also list a binary flag that contains additional information about the candidate binary. For this section, we most importantly list if a binary is in a high contamination region such that the clean sample can be recovered. \textcolor{black}{We want to clarify that this table contains \textit{all} candidate binaries within 200 pc. To get the likely binaries, it is recommended to select binaries from Table \ref{tab:candidate_binaries} with $L < 0.00193$. For the subsequent analyses in this paper, when we refer to likely binaries we are referring to candidate binaries with this $L < 0.00193$ cut applied. Additionally, to get the clean sample where binaries in high contamination regions are removed, additionally select binaries with $C_{ideal} < 0.1$. For the subsequent analyses in this paper, the clean sample will be candidate binaries with $L < 0.00193$ and $C_{ideal} < 0.1$. If users would like to make their own quality cuts, all values of the contamination factor, $L$, and the contamination rate per healpix bin calculated for an $L_{cut}$ that selects all candidates, $C_{all}$, are provided in Table \ref{tab:candidate_binaries}.}

\begin{sidewaystable*}
\tiny
	\centering
	\caption{\textcolor{black}{All }Gaia eDR3 \textit{red}{candidate binaries} within 200 pc. In the below columns, the subscript 1 identifies the primary star, which is the star used to measure the $G-J$ excess in the probability distributions, and the subscript 2 identifies the neighboring secondary star. \textcolor{black}{This table includes \textit{all} candidate binaries. To get the likely binaries, it is recommended to select binaries with $L < 0.00193$. Additionally, to get the clean sample where likely binaries in high contamination regions are removed, select binaries with $L < 0.00193$ and $C_{ideal} < 0.1$. If users would like to make their own quality cuts, all values of the contamination factor, $L$, and the contamination rate per healpix bin calculated for an $L_{cut}$ that selects all candidates, $C_{all}$, are provided in the table below for all candidate binaries.}}
	\label{tab:candidate_binaries}
	\begin{tabular}{ccccccccccccccccc} 
		\hline
		Gaia eDR3 ID$_1$ & Gaia eDR3 ID$_2$ & $\alpha_1$ & $\delta_1$ & $\theta$ & $\pi_1$ & $J_1$ & $G_1$ & $BP_1$ & $RP_1$ & $G_2$ & $BP_2$ & $RP_2$ & $L$ & $C_{ideal}$ & $C_{all}$ & Binary Flag$^\dagger$ \\
		  & & [deg] & [deg] & ["] & [mas] & [mag] & [mag] & [mag] & [mag] & [mag] & [mag] & [mag] & & & & \\
		\hline
		2846325099652759680 &2846325099651932416  & 0.44429056961&  20.13333491082   &  0.7110  & 6.5545328550801205  &   9.965& 11.739593 &11.904744 &10.721714 &12.666404  &\nodata & \nodata&  0.000000& $-$0.249 & 0.057 & 9 \\
		2846391551386740992 &2846391551385854592 &  0.28305737941&  20.36364725895  &   0.5034  & 5.5107728050205065   & 16.738 &19.348417&19.605074& 17.730440 &19.507917  &\nodata &  17.753252 & 0.026522& $-$0.249&  0.057  &9\\
		2846394746841506560 &2846394746842403072 & 0.12007216481&  20.44873497076   &  0.6657  & 5.0467249613545038 &   15.787 &18.787980 &19.693420&17.346070& 19.928154  &\nodata & \nodata&  0.000770& $-$0.249 & 0.057 & 9 \\
		2846518407540379008 &2846518403244802304 &  0.23100930282 & 20.91763780491  &   0.9051  & 5.6103156188371113  &  13.339& 14.995512& 14.804933& 13.908501& 15.322645  &\nodata & \nodata&  0.000000& $-$0.249 & 0.057  &9 \\
		2846731094321228032 &2846731094320213376 &  0.02161906395 & 21.32031168184 &    0.6454  & 7.6188382788615163 &   11.986 &14.938788 &16.003300& 13.557371 &16.407385 &\nodata & \nodata&  0.000000& $-$0.249&  0.057 & 9\\
		2847246176863467648& 2847246176864570112  & 0.53453378103 & 22.35137729473   &  1.8568  & 5.2753888179350445  &  14.518& 17.663988& 19.180538 &16.325012& 19.211802  &\nodata & \nodata& 0.000259& $-$0.545 & $-$0.127&  3\\
		2847357575430420992 &2847357575430421120  & 1.12321296810 & 22.73517731748  &   2.1465  & 9.7243137010788505 &    8.953 &10.275372& 10.634806 & 9.741213 &12.771252& 13.169403 &11.688901&  0.000001 &$-$0.545&  $-$0.127 & 3 \\
		2847376885603489536 &2847376885603489408&   1.68221159003 & 22.95113092277  &   2.3833  & 6.1633957969951823   & 10.735 &13.181674 &13.887457& 12.258460 &13.465112 &14.206071 &12.590742 & 0.000001 &$-$0.545 &$-$0.127 & 3\\
		392484114490982656  &392484114492364416  & 5.49956488278 & 47.51785345138    & 2.0964 & 13.6232947400285642 &   11.319& 14.106323 &15.542462 &12.929996 &19.205818  &\nodata & \nodata&0.155536& $-$0.171  &0.586 & 3\\
		392497751011995008 & 392497755307030528  & 5.38306248684 & 47.83016817900 &    1.2025  & 6.0039775674163796  &  10.674& 12.165501 &12.567431& 11.458676 &14.635649&\nodata & \nodata&  2.641787& $-$0.857 & 0.426&  5 \\
		\hline
	\end{tabular}
\\
  NOTE -- This table is published in its entirety in a machine-readable format. A portion is shown here for guidance regarding its form and content.
  \\
  $^\dagger$Binary flag with bit information: (1) Binary in low contamination region (\textcolor{black}{$<10\%$}) (2) Binary identified by \citet{elbadry_2021}, (4) Binary identified as triple system with \citet{elbadry_2021} binary, (8) Secondary has no parallax measurement in Gaia eDR3, (16) Binary physical separation $<30$ AU with parallax error $<5\%$ and (32) Binary physical separation $<30$ AU with parallax error $>5\%$.
\end{sidewaystable*}

\section{Discussion}\label{sec:discuss4}

\subsection{Comparison of \textcolor{black}{Likely} Binaries to Known Binaries in Gaia eDR3}

In an attempt to validate some \textcolor{black}{likely} binaries in our sample, we compare our sample to the binary catalog from \citet{elbadry_2021}. \textcolor{black}{Again, for this analysis we are only considering likely binaries with $L < 0.00193$.}

Out of the \textcolor{black}{68,725}(\textcolor{black}{50,230}) total(clean; \textcolor{black}{defined as $C_{ideal} < 0.1$}) \textcolor{black}{subset of likely} binaries identified in the present work, we find that \textcolor{black}{24,899(22,963)} of the \textcolor{black}{likely binaries} are already listed in the \citet{elbadry_2021} catalog. These binaries are indicated in Table \ref{tab:candidate_binaries} by the corresponding binary flag. It is easy to understand why over half of our \textcolor{black}{likely binaries} were not identified in the construction of the \citet{elbadry_2021} catalog; of the remaining \textcolor{black}{43,826(27,267)} \textcolor{black}{likely binaries}, \textcolor{black}{37,075(22,824)} have neighbors with no parallax measurements, \textcolor{black}{3,851(1,880)} have neighbors with \texttt{parallax\_over\_error} $<5$, and \textcolor{black}{230(91)} have neighbors with \texttt{parallax\_error} $>2$, where each of these groups would violate the initial sample selection by \citet{elbadry_2021}. This demonstrates one advantage of the present \textcolor{black}{likely binary} list, which tends to include companions with shorter angular separations and fainter magnitudes, from which reliable astrometric solutions are not yet available. If the \textcolor{black}{likely} companions can be confirmed by follow-up observations, this would provide reliable distances for them, and would add thousands of low-separation binaries to the Solar Neighborhood census. There are a total of 41,519 stars in the \citet{elbadry_2021} catalog with angular separations $<2.5$" and $d<200$ pc, so if all \textcolor{black}{likely} binaries in this study were confirmed this would lead to a \textcolor{black}{$\sim$106\%} increase in the number of known low angular separation binaries in the Solar Neighborhood from Gaia eDR3.

Another matter of interest resulting from our improved list of \textcolor{black}{likely binaries} is the identification of new visual triples. Specifically, by identifying \textcolor{black}{likely} binaries in our sample that also happen to be one of the components of a wider \citet{elbadry_2021} binary with a listed separation $>2.5$", we can curate a list of newly identified visual components in these now \textcolor{black}{likely} triple systems. We identify \textcolor{black}{1,232(917)} such systems in our list of \textcolor{black}{likely} binaries. These candidate triple systems are indicated in Table \ref{tab:candidate_binaries} by the corresponding binary flag.

\begin{figure}[!t]
	\centering
	\includegraphics[width=0.5\textwidth]{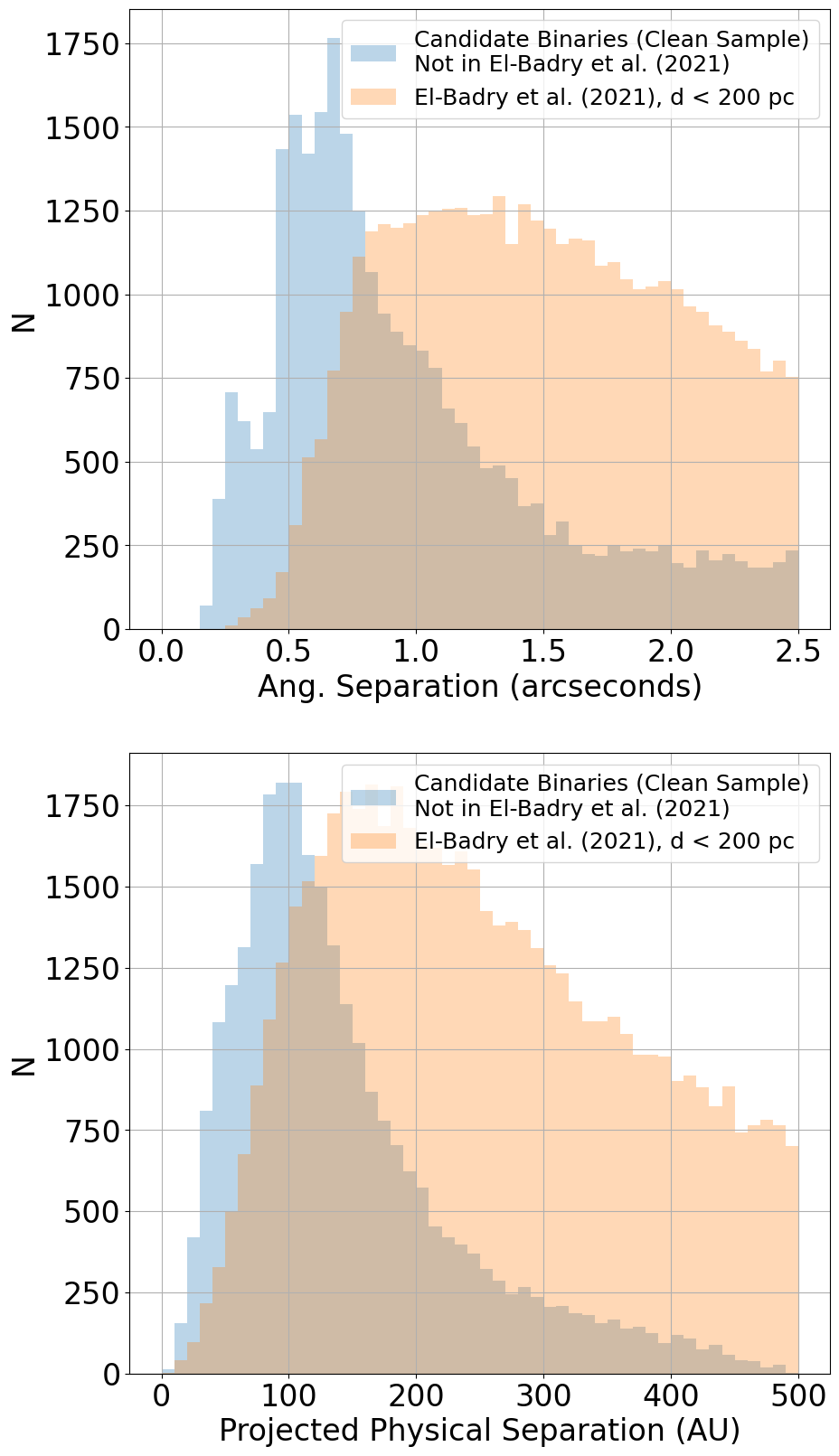}
	\caption{Distributions of angular separation (top panel) and projected physical separation (bottom panel) for the candidate binaries with \textcolor{black}{$L < 0.00193$} and not in a high contamination region, and that were not previously found in the \citet{elbadry_2021} catalog (blue histograms). All the binaries in the \citet{elbadry_2021} catalog within 200 pc are also shown for comparison (orange histograms). For the projected physical separation for the \textcolor{black}{likely} binaries, the parallax of the Gaia eDR3 200 pc source was used to calculate projected physical separation, as many of the companions do not have parallax measurements.}
	\label{fig:sep_comp_el_badry}
\end{figure}

Also, by comparing the distribution of angular separations for the new \textcolor{black}{likely} binaries to that of the known binaries in the \citet{elbadry_2021} catalog, we find that our method is significantly expanding the search of companions to smaller angular and physical separations. The top panel of Figure \ref{fig:sep_comp_el_badry} shows the angular separation distributions for our \textcolor{black}{likely} binaries not identified by \citet{elbadry_2021} (blue histograms) and for all binaries in the \citet{elbadry_2021} catalog within 200 pc (orange histograms). This demonstrates that our sample significantly increases Gaia \textcolor{black}{likely} binaries with small angular separations, notably for $\theta<0.7"$. Additionally, when comparing these two subsets in terms of projected physical separation (Figure \ref{fig:sep_comp_el_badry}, bottom panel), our new \textcolor{black}{likely binaries} have the potential to nearly triple the number of resolved nearby binaries in Gaia eDR3 with projected separations $<100$ AU.

\begin{figure}[!t]
	\centering
	\includegraphics[width=0.5\textwidth]{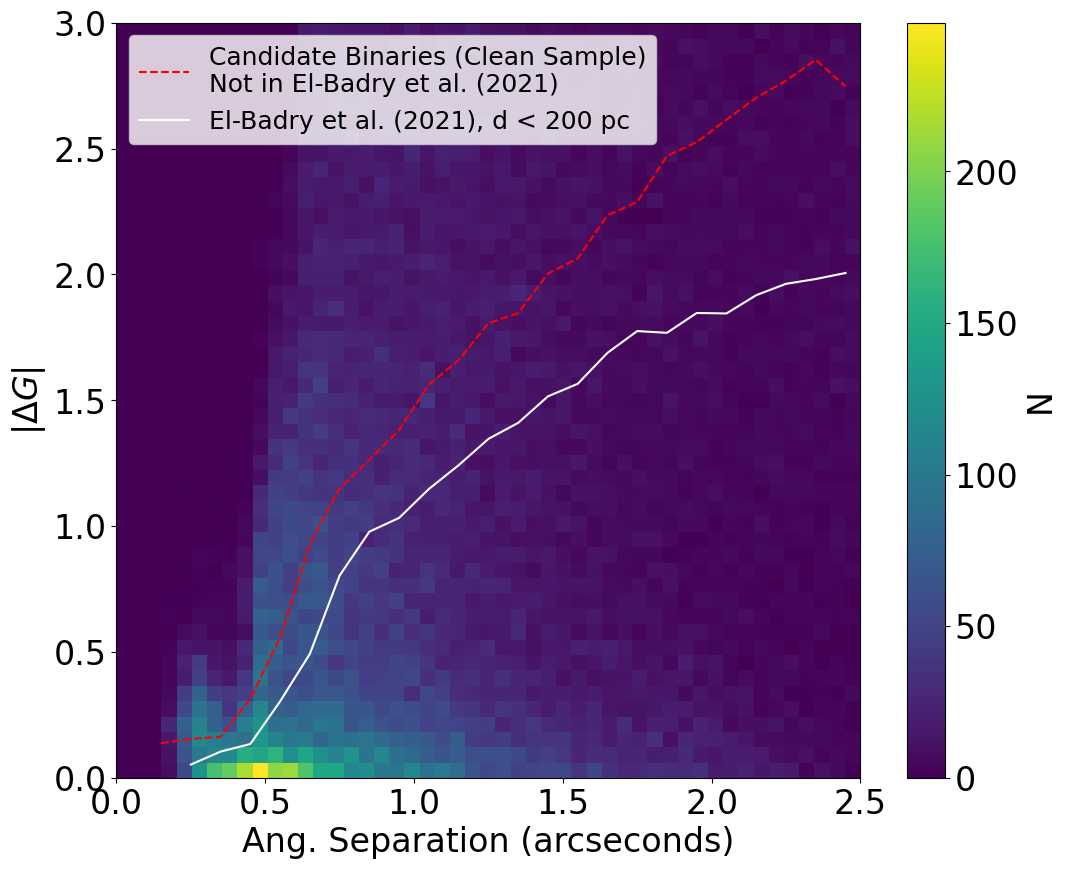}
	\caption{Density distribution of $|\Delta G|$ as a function of angular separation for the candidate binaries with \textcolor{black}{$L < 0.00193$} and not in a high contamination region. Overlayed is the 1$\sigma$ \textcolor{black}{scatter} of $|\Delta G|$ \textcolor{black}{values in bins of} angular separation \textcolor{black}{of width} 0.1" for candidate binaries with \textcolor{black}{$L < 0.00193$}, not in a high contamination region and not found in the \citet{elbadry_2021} catalog (red dashed line), and all binaries in the \citet{elbadry_2021} catalog within 200 pc (solid white line). \textcolor{black}{This demonstrates that a larger portion of the new pairs in our catalog have larger $|\Delta G|$ values than in the \citet{elbadry_2021} catalog.}}
	\label{fig:delta_G_comp_el_badry}
\end{figure}

Another improvement is in expanding the range of the mass ratio of the \textcolor{black}{likely} binaries. Qualitatively, this can be probed by the $|\Delta G|$ values for the components of the system, where larger values indicate a smaller value of mass ratio, $q$. Figure \ref{fig:delta_G_comp_el_badry} shows the distribution $|\Delta G|$ as a function of angular separation for the candidate binaries with \textcolor{black}{$L < 0.00193$} and not in a high contamination region, where overlayed on this distribution is the 1$\sigma$ of $|\Delta G|$ as a function of angular separation in bins of 0.1" for \textcolor{black}{likely} binaries not found in the \citet{elbadry_2021} catalog (red dashed line) and all binaries in the \citet{elbadry_2021} catalog within 200 pc (solid white line). For all angular separations we find that we push to lower mass ratios compared to \citet{elbadry_2021}. This indicates that our sample of new \textcolor{black}{likely} binaries is not just incremental: it also expands the range (and statistics) of mass-ratio distribution for systems with separations $<100$ AU.

\subsection{Determining Orbits with Followup Observations for New \textcolor{black}{Likely Binaries}}

As mentioned above, our sample of \textcolor{black}{likely} binaries significantly expands on the number of short-separation binaries in Gaia eDR3. One critical application of such systems is the ability to derive their gravitational masses through astrometric monitoring. Such monitoring has become easier with recent advances in speckle imagine which allow for precise positioning of small separation stars with fairly little observational overhead \citep{speckle_letter}. 

With this in mind, in our proposed sample we have numerous \textcolor{black}{binaries} that would be ideal for astrometric monitoring over the coming decades. From the \textcolor{black}{likely binaries} in Table \ref{tab:candidate_binaries} not in high contamination zones and not previously identified by \citet{elbadry_2021}, we find \textcolor{black}{420(221)} stars with projected physical separations $20<s<30$ AU (and with parallax errors $<5\%$), \textcolor{black}{156(121)} with $10<s<20$ AU and \textcolor{black}{14(14)} with $s<10$ AU. These short-separation binaries are indicated in Table \ref{tab:candidate_binaries} by the corresponding binary flag. This is a large increase compared to known Gaia eDR3 binaries from \citet{elbadry_2021} within 200 pc, where they had found 96(93) stars with $20<s<30$ AU \textcolor{black}{(and with parallax errors $<5\%$)}, 40(40) with $10<s<20$ AU and 2(2) with $s<10$ AU. All \textcolor{black}{696} systems not in high contamination regions and with separations $<30$ AU found in the present study are listed in Table \ref{tab:low_sep_binaries}. Additionally, we note in this table all \textcolor{black}{likely} binaries that were previously identified in the Washington Visual Double Star Catalog \citep[WDS;][]{WDS} by indicating their WDS name \textcolor{black}{(up-to-date with WDS as of September 12th 2023)}. Only \textcolor{black}{142} of the \textcolor{black}{696} low-separation \textcolor{black}{($s < 30$ AU)} \textcolor{black}{systems} have been cataloged in the WDS thus far.

\begin{figure*}[!t]
	\centering
	\includegraphics[width=0.8\textwidth]{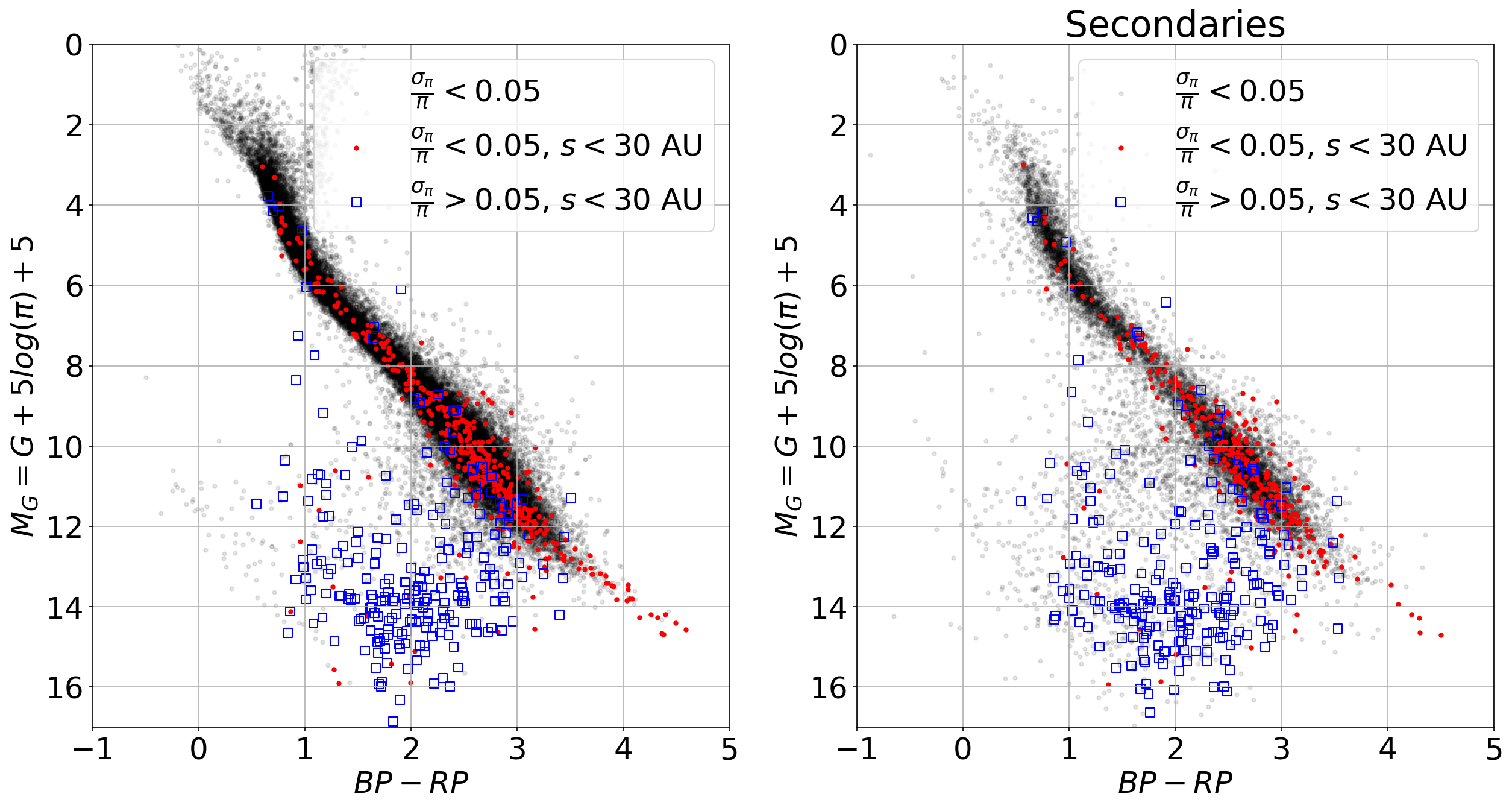}
	\caption{HR diagram for a subset of the candidate binaries with \textcolor{black}{$L < 0.00193$} and not in high contamination regions, where primaries are shown in the left panel and secondaries in the right panel. In both panels, the parallax of the primary is used to calculate the absolute magnitude. The black data points are for \textcolor{black}{likely} binary primaries with parallax error $<5\%$, open blue squares are for \textcolor{black}{likely} binaries not identified by \citet{elbadry_2021} with $s<30$ AU and parallax error $>5\%$, and red closed circles are for \textcolor{black}{likely} binaries not identified by \citet{elbadry_2021} with $s<30$ AU and parallax error $<5\%$.}
	\label{fig:hr_diag_low_sep}
\end{figure*}

The HR diagram distributions of \textcolor{black}{likely} binaries with low-separations \textcolor{black}{($s < 30$ AU)} found in this study are shown in Figure \ref{fig:hr_diag_low_sep}, where we find that the majority of the low-separation binaries are stars of K and M spectral types. Additionally, some low separation \textcolor{black}{binaries} show significant overluminosities in the HR diagram, which could be an indication of additional unresolved stars in one of the components. We have begun an initial speckle campaign of some these \textcolor{black}{likely} binaries that have not been observed with high resolution imaging yet and these observations will be discussed in future paper.

Overall, it should be understood that the orbital periods of most of these systems may be $>100$ years, which means long-term observations will be required to even estimate a preliminary orbit in a few decades. Using SIMBAD \citep{simbad}, we have identified a number of systems with past observations via high resolution imaging or astrometric anomaly\textcolor{black}{, where this search is up to date as of September 12th 2023}. \textcolor{black}{Here, astrometric anomaly detections rely on the difference between the center of light for a binary system over time and the expected motion of a single star \citep{Penoyre2020}, such that deviations from an astrometric solution indicate probable orbital motion for barely resolved sources.} We identify these systems in Table \ref{tab:low_sep_binaries}, and list the three most recent studies that have imaged these systems or identified them via astrometric anomaly. For \textcolor{black}{4} out of the \textcolor{black}{15} shortest separation systems ($s<10$ AU), however, there are still no observations with high resolution imaging. For these systems, the final data release of Gaia, which will have all epoch data for all stars, may be used to determine preliminary orbits, potentially leaving just a few more epochs of data to get reliable orbital determinations. Regardless, these new systems, should be included in follow-up campaigns to 1) confirm the binary status of the system and 2) begin to map out their orbits.

\begin{table*}
	\tiny
	\centering
	\caption{Gaia eDR3 stars within 200 pc with \textcolor{black}{$L < 0.00193$}, not in a high contamination regions and having a projected physical separation $<30$ AU. The subscript 1 identifies the primary star, which is the star used to measure the $G-J$ excess in the probability distributions, and the subscript 2 identifies the neighboring secondary star. The final column of the table provides the three most recent studies that have imaged, identified or determined orbits for these systems, if any exist.}
	\label{tab:low_sep_binaries}
	\begin{tabular}{ccccccccccc} 
		\hline
		Gaia eDR3 ID$_1$ & Gaia eDR3 ID$_2$ & WDS &$\alpha_1$ & $\delta_1$ & $s$ & $G_1$ & $BP_1$ & $RP_1$ & $G_2$ & Citations$^\dagger$ \\
		& & & [deg] & [deg] & [AU] & [mag] & [mag] & [mag] & [mag] & \\
		\hline
		2862257023139546752 &2862257023138466048  & \nodata &  5.97558816300 & 31.33549471509& 29.60 &18.367035& 19.365164& 16.344255& 18.410372     & \nodata \\       
		2360176583186317184& 2360176583187424000   & \nodata &   5.09812972263 &$-$23.76855271948 &12.22& 17.006517 &19.077362 &14.816156 &17.084696    & \nodata \\
		2368229058456070656& 2368229062751642112    & \nodata &     4.63882020362 &$-$16.45080356495 &28.66& 14.560919 &15.067697 &12.738357 &14.582545   & \nodata \\
		2880681298968094336& 2880681298966894720  & \nodata &   1.67367154586  &38.06778981192 &25.61 &17.524828& 17.217081 &16.238302& 17.630812   & \nodata \\
		2877469591144616320 &2877469591142038400   & \nodata &  3.00162605288 & 37.17893035300 &21.62 &19.507418 &19.934586& 17.662619& 19.499413   & \nodata \\
		2416052011764454144& 2416052011763252992   & \nodata &   0.30642590584 &$-$15.44086795687 &22.67& 18.820293& 19.263636& 17.138237&18.649157  & \nodata \\
		2417069815933263360 &2417069815934357376& WDS J00162-1435&   4.04520064244& $-$14.59194690502 &14.79& 12.226642 &13.040962& 10.501659& 12.551744 &2 \\         
		2417948085206509952&2417948085206852224    & \nodata &   2.50715937956 &$-$13.91103631239 &14.39& 13.829072 &14.552967& 11.972415& 13.859048 & \nodata \\ 
		2793964637951353088& 2793964637950130304    & \nodata &    6.16188283670 & 17.45297866140& 28.99& 19.393719 &19.446209 &17.977188 &19.601963   & \nodata \\
		382041296646363904 & 382041296644747008  & \nodata &     6.21236790988  &41.68403511126 &30.00 &18.486490& 19.635220 &16.386400 &18.577877  & \nodata \\
		\hline
	\end{tabular}
	\\
	NOTE -- This table is published in its entirety in a machine-readable format. A portion is shown here for guidance regarding its form and content.
	\\
	\textcolor{black}{$^\dagger$Citation flags correspond to following studies: (1) \citet{1972AJ.....77..878W} (2) \citet{1975ApJS...29..315H} (3) \citet{1980ApJS...44..111H} (4) \citet{1982AAS...50...49C} (5) \citet{1985AAS...60..241C} (6) \citet{1987AJ.....93..688M} (7) \citet{1987ApJS...65..161H} (8) \citet{1993AAS..100..305C} (9) \citet{1993AJ....106..352H} (10) \citet{1994AAS..106..377C} (11) \citet{1994RMxAA..28...43P} (12) \citet{1996AJ....111..393A} (13) \citet{1997AAS..126....1G} (14) \citet{1997AJ....114.1623F} (15) \citet{1998ApJS..117..587H} (16) \citet{2000AAS..145...67M} (17) \citet{2001AJ....121.3259M} (18) \citet{2002AA...384..180F} (19) \citet{2004AA...422.1023S} (20) \citet{2008AJ....135.1334H} (21) \citet{2008AJ....135.1803D} (22) \citet{2008MNRAS.384..150L} (23) \citet{2009AJ....138..813H} (24) \citet{2010AA...520A..54B} (25) \citet{2010AJ....139..743T} (26) \citet{2010ApJS..190....1R} (27) \citet{2010RMxAA..46..245O} (28) \citet{2011AJ....141...45H} (29) \citet{2011AJ....142...46M} (30) \citet{2011RMxAA..47..211O} (31) \citet{2012AJ....143...10H} (32) \citet{2012AJ....143...42H} (33) \citet{2012ApJ...754...44J} (34) \citet{2012MNRAS.421.2498G} (35) \citet{2013AJ....146...56M} (36) \citet{2013MNRAS.429..859J} (37) \citet{2014ApJ...789..102J} (38) \citet{2015AJ....149..151H} (39) \citet{2015AJ....150...50T} (40) \citet{2015AJ....150..151H} (41) \citet{2015MNRAS.449.2618W} (42) \citet{2016AJ....152..216G} (43) \citet{2017AA...599A..70J} (44) \citet{2017AJ....153..212H} (45) \citet{2018AA...619A..81H} (46) \citet{2018AJ....155..215M} (47) \citet{2019AA...623A..72K} (48) \citet{2019AJ....157..211M} (49) \citet{2019AJ....157..216W} (50) \citet{2019AJ....158..167T} (51) \citet{2019ApJ...877...60B} (52) \citet{2019MNRAS.482.4096D} (53) \citet{2020AJ....159..139L} (54) \citet{2020AJ....159..233H} (55) \citet{2020AJ....160....7T} (56) \citet{2020AJ....160..120J} (57) \citet{2020AJ....160..215V} (58) \citet{2021AJ....161..295H} (59) \citet{2021AJ....162...53M} (60) \citet{2021AJ....162..102S} (61) \citet{2021AJ....162..156M} (62) \citet{2021ApJS..254...42B} (63) \citet{2022AA...666A..16C} (64) \citet{2022AJ....163..178V} (65) \citet{2022AJ....163..200S} (66) \citet{2023AJ....165..193W}.}
\end{table*}

\subsection{Determining Physical Separation Distributions and Binary Fractions}

Another application of the sample of close visual binaries assembled here is to constrain the multiplicity fraction and distribution of orbital separations notably for low-mass stars in the Solar Neighborhood. We provide a preliminary assessment in this section.

First, we determine the expected completeness of our binary star sample based on the sample selection cuts for the 200 pc subset (see Section \ref{sec:data4}). \citet{elbadry2018} demonstrated how one can use the estimated distribution of field stars around a sample of stars that match your proposed sample selection criteria to estimate the completeness of the sample as a function of angular separation and $|\Delta G|$. In this method, a dense region on the sky is queried, and all stars that match the proposed sample selection criteria are identified. Then, all field stars out to some angular separation limit are also identified. Using the resulting separations to these stars (which include both field stars and binary companions), the distribution of angular separation is plotted for various bins of $|\Delta G|$. Then, a linear trend is fit to the larger angular separation portion of these distributions. Here we fit the linear trend for parts of the distribution where $\theta >$2.5". The linear trend extrapolated to smaller angular separations normally represents the number of stars one would expect to detect if there were no sensitivity issues for different values of $|\Delta G|$ in the Gaia catalog. \textcolor{black}{So, if there are any deficits compared to the expected distribution, this indicates that the sample is not complete in this smaller angular separation regime.}

\begin{figure*}[!t]
	\centering
	\includegraphics[width=0.45\textwidth]{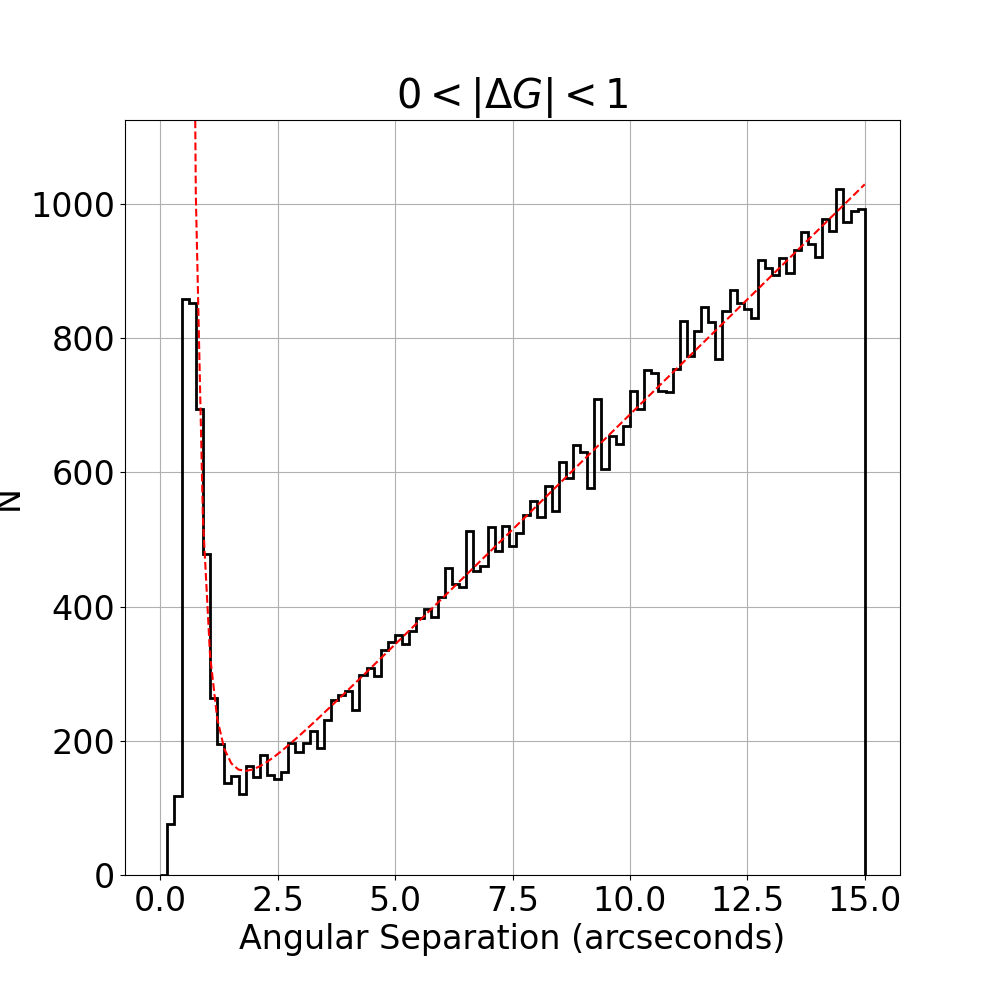}
	\includegraphics[width=0.45\textwidth]{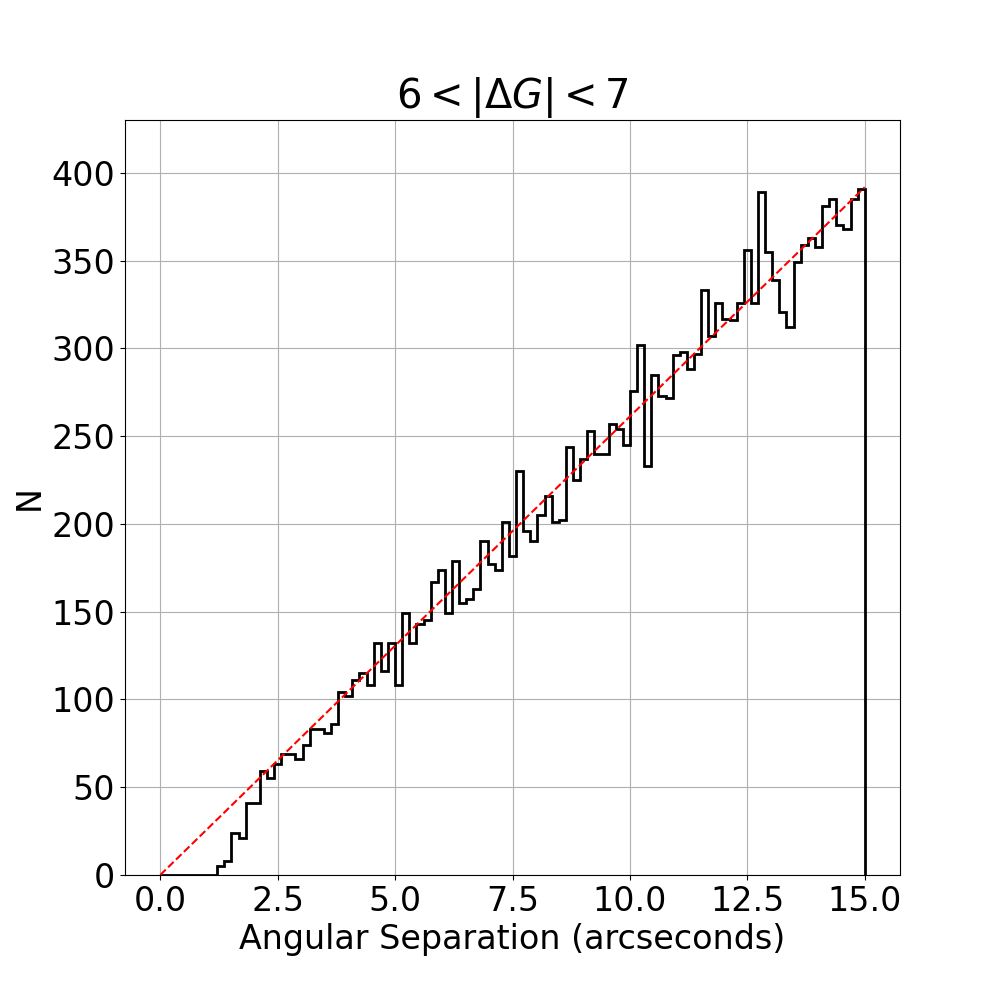}
	\caption{Selection of Gaia stars that result from querying stars in our 200 pc sample within $10^\circ$ of $(l,b) = 330^\circ, -4^\circ)$. The left panel shows the observed distribution (black line) for pairs of stars with $0 < |\Delta G| < 1$ and the right panel for pairs with  $6 < |\Delta G| < 7$. The red dashed lines shows the fit to the distributions, where the left panel fits a line in the outer region ($\theta >$2.5") and a line plus a power law in the inner region (0.7"$<\theta <$2.5"), while the right panel just fits a line to the outer region.}
	\label{fig:sep_complete_test}
\end{figure*}

This comes with a caveat: while it is true that chance alignments dominate for larger $|\Delta G|$ values with our sample selection, this is not actually true for smaller values of $|\Delta G|$, as is shown in Figure \ref{fig:sep_complete_test}. For pairs with $|\Delta G|$ < 1 we actually see a large excess of stars above the linear trend for small angular separations\textcolor{black}{, which is due to true binaries that exist in the selection and dominate the small separations for small magnitude differences}. This means to account for the observed distribution, we have to fit an additional component in this inner region. So, for instances where in the inner region (0.7"$<\theta <$2.5") we find that there are counts greater than the extrapolated linear trend, we fit an additional power law in this region (after subtracting off the linear trend fit in the outer region). The sum of these two components \textcolor{black}{is then used to describe the expected distribution of stars and, again, any deficits from this relationship at small angular separations indicate completeness issues}. The left panel of Figure \ref{fig:sep_complete_test} demonstrates that this better explains the observed distribution and that we similarly still observe a deficit of stars at smaller angular separations.

Next, the observed distribution is divided by these fits so we can then find the completeness level as a function of angular separation for each bin of $|\Delta G|$. We then fit the resulting distribution with the relationship of expected fraction detected as a function of angular separation, described in \citet{elbadry2018} as:
\begin{equation}\label{eq:fraction_eq}
	f_{|\Delta G|}(\theta) = \frac{1}{1 + (\theta / \theta_0) ^ {-\beta}}
\end{equation}
Additionally, when fitting the above equation, \citet{elbadry2018} found that $\theta_0$ generally was linearly related to $|\Delta G|$ and that the median value of $\beta$ could describe all fits. We will also follow this procedure for our final fits.

\begin{figure}[!t]
	\centering
	\includegraphics[width=0.5\textwidth]{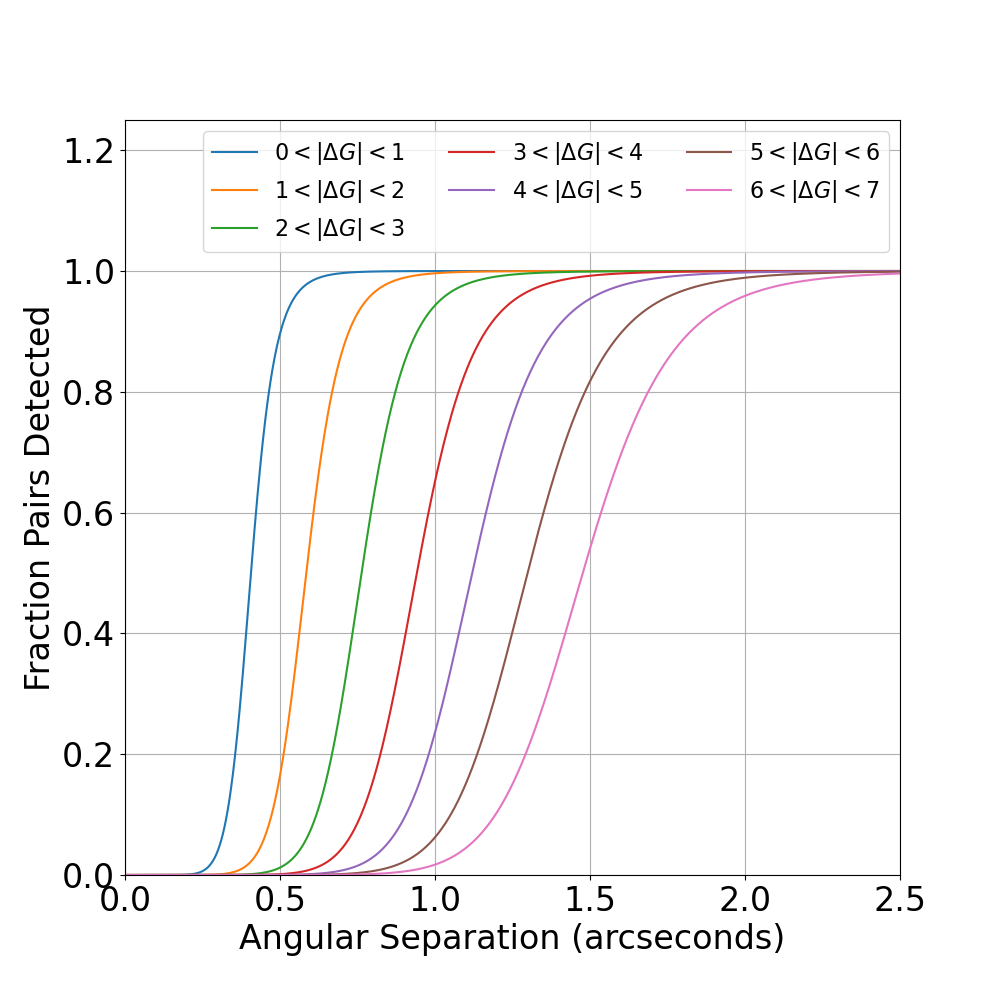}
	\caption{The expected fraction of stars detected as a function of angular separation for various bins of $|\Delta G|$ given the sample selection for this study (outline in Section \ref{sec:data4}).}
	\label{fig:fraction_detected}
\end{figure}

\textcolor{black}{While we cannot directly use the results from \citet{elbadry2018} as their sample selection from \textit{Gaia} is different from ours, we will repeat the above procedure with the sample selection used in this paper (see Section \ref{sec:data4}) for stars} within $10^\circ$ of $(l,b) = (330^\circ, -4^\circ)$. \textcolor{black}{With this sample, we then fit either the linear or linear$+$power-law distributions (Figure \ref{fig:sep_complete_test}) for the various magnitude differences of the stars in the sample. After dividing these fits by the observed distribution,} using eq. \ref{eq:fraction_eq} we find that $\theta_0(|\Delta G|) =  0.178 |\Delta G| + 0.317$ and that the median value of $\beta$ is 10.383. Using these fits, the resulting estimated fraction of stars detected with our sample selection criteria per angular separation and magnitude difference are shown in Figure \ref{fig:fraction_detected}. These fits indicate that our sample should be 95\% complete for $\theta \gtrapprox 0.776$" and $| \Delta G| < 2$. So, when examining binary separation distributions we will only consider \textcolor{black}{likely} binaries that fit this completeness criteria.

\begin{figure*}[!t]
	\centering
	\includegraphics[width=0.66\textwidth]{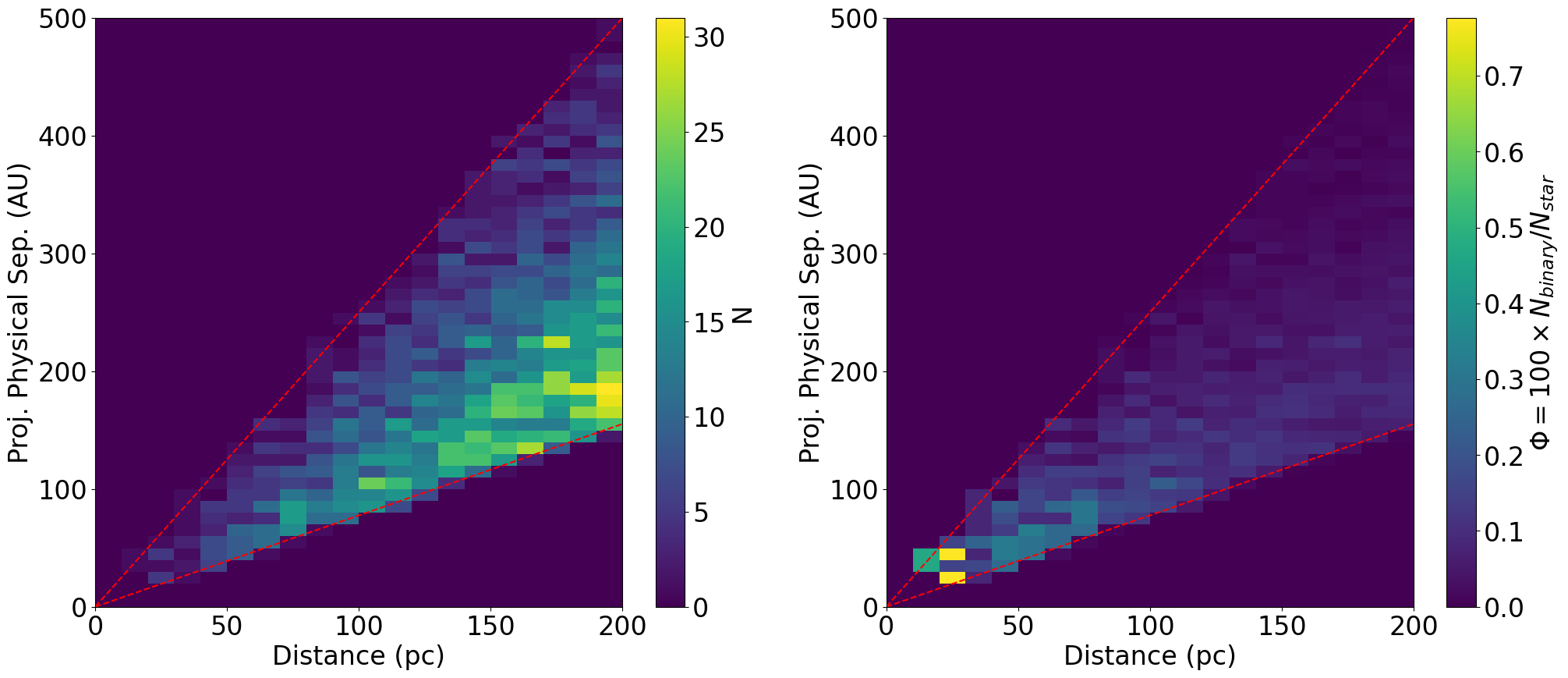}
	\includegraphics[width=0.3\textwidth]{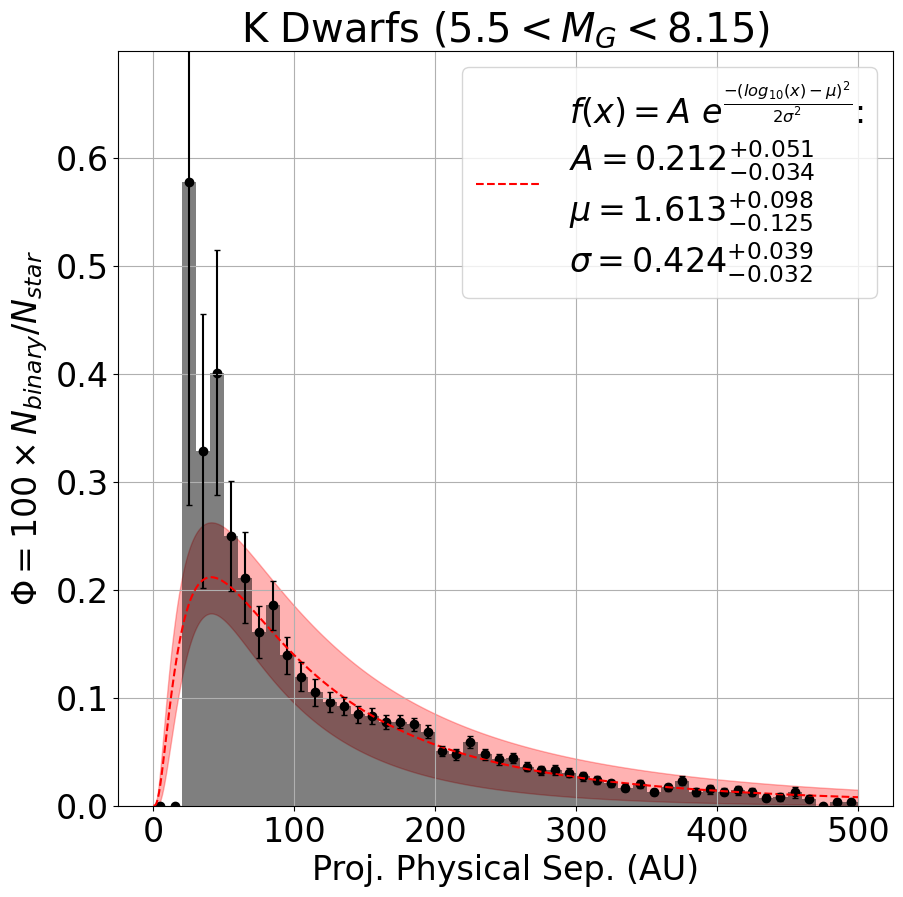}
	\caption{\textbf{Left Panel:} Number density of projected physical separation versus distance for \textcolor{black}{likely} binaries with $\theta > 0.776$", $| \Delta G| < 2$ and $5.5 < M_G < 8.15$. The red dashed lines indicate the minimum ($\theta = 0.4$") and maximum ($\theta = 2.5$") angular separation limits imposed. \textbf{Middle Panel:} Number density plot shown in the left panel where each column has been normalized by the volume probed by each bin. Additionally, each bin has been arbitrarily scaled by 100 for readability. \textbf{Right Panel:} Distribution of projected physical separations (black histogram) for the sample in the left panel. This distribution has been calculated by taking the average in each row in the middle panel where $100 \times N_{binary} / N_{star} > 0$. The distribution was bootstrapped over 1000 iterations to produce the uncertainties shown as the black error bars. The distribution is fit with a log-normal function of the form: $f(x) = A \ e^{\frac{-(log_{10}(x) - \mu)^2}{2 \sigma^2}}$\textcolor{black}{, where $x =s$ and $s$ is the linear, projected physical separation in AU}. The posterior distribution for the model parameters are probed using \textit{emcee} and the resulting 16th, 50th and 84th percentile of the distributions are shown in the legend of the figure.}
	\label{fig:ex_phys_sep}
\end{figure*}

To determine binary physical separation distributions, we still must consider the completeness issues that are a result of the minimum and maximum angular separations imposed by the method used in this study. This completeness issue is illustrated in the left panel of Figure \ref{fig:ex_phys_sep}, where it is clear that for certain regions in physical separation versus distance, we are missing data due to the imposed angular separation limits (red dashed lines in Figure \ref{fig:ex_phys_sep}). This means that simply plotting the projected physical separation distribution would lead to an under-counting of binaries at small projected physical separations. One way to mitigate this is to normalize this distribution by the number of stars in the overall \textit{Gaia}-2MASS 200 pc sample in each column. This normalization will account for any completeness issues that may be a function of distance for each spectral type range probed allowing us to find the number of binaries per star per projected physical separation bin. This results in the distribution in the middle panel of Figure \ref{fig:ex_phys_sep}. With the distribution properly normalized, we can find the expected physical separation distribution by simply taking the average of each row in the normalized distribution for bins where $100 \times N_{binary} / N_{star} > 0$. To deal with the low number statistics of the bins at lower projected physical separation, we bootstrap these averages for 1000 iterations of the distribution. The average projected physical separation distribution resulting from this bootstrapping procedure is shown in the right panel of Figure \ref{fig:ex_phys_sep}, where the average is shown as the black histograms and the $1\sigma$ from this average is shown with the black error bars. Finally, we fit the resulting distribution with a log-normal distribution of the form:
\begin{equation}
	f(x) = A \ e^{\frac{-(log_{10}(x) - \mu)^2}{2 \sigma^2}}
\end{equation}
\textcolor{black}{In the above $x =s$ and $s$ is the linear, projected physical separation in AU.} The posterior distribution for the model parameters are probed using \textit{emcee} \citep{emcee} and the resulting 16th, 50th and 84th percentile of the distributions are shown in the legend of the right panel of Figure \ref{fig:ex_phys_sep}.

\textcolor{black}{We perform this procedure separately for stars of K and early M spectral types.} It should be noted that not all spectral type ranges are expected to be volume complete for the distance probed ($d<200$ pc), but by dividing the distribution by the total number of stars in the overall \textit{Gaia}-2MASS 200 pc sample for each spectral type and distance bin, we can account for such incompleteness issues. The resulting \textcolor{black}{log-normal} fits for \textcolor{black}{these two spectral types} are shown in Table \ref{tab:binary_fractions}.

\begin{table*}
	\scriptsize
	\centering
	\caption{Summary of fits to physical separation distributions and resulting estimated binary fractions for various spectral types of the primary star, as selected with cuts in $M_G$. Fits to the physical separation distributions are done using a \textcolor{black}{log-normal function of the form: $f(x) = A \ e^{\frac{-(log_{10}(x) - \mu)^2}{2 \sigma^2}}$, where $x =s$ and $s$ is the linear, projected physical separation in AU.} The posterior distribution for the model parameters are probed using \textit{emcee} and the resulting 16th, 50th and 84th percentile of the distributions are shown. The final three columns of the table shows the binary fraction for the sub-sample of \textcolor{black}{likely} binaries using eq. \ref{eq:bin_frac} for various ranges of $[s_{min}, s_{max}]$.}
	\label{tab:binary_fractions}
	\begin{tabular}{lccccccc} 
		\hline
		Primary Spectral Type & $M_G$ Cut &  $A$ & $10^\mu$ & $\sigma$ & $f$ ($s\in[30, 500]$ AU) & $f$ ($s\in[30, 200]$ AU) & $f$ ($s\in[200, 500]$ AU) \\
		 &   &   & [AU] &  & & &  \\
		\hline
		\hline		
		K Dwarfs & ($5.5 < M_G < 8.15$) & $0.212_{-0.034}^{+0.051}$ & $40.996_{-11.780}^{+9.208}$ & $0.424_{-0.032}^{+0.039}$ & $0.298 \pm 0.125$ & $0.218 \pm 0.072$ & $0.079 \pm 0.055$ \\
		M Dwarfs & ($8.15 < M_G < 13$) & $0.247_{-0.021}^{+0.025}$ & $37.965_{-4.472}^{+3.884}$ & $0.393_{-0.014}^{+0.016}$ & $0.275 \pm 0.051$ & $0.224 \pm 0.034$ & $0.050 \pm 0.018$ \\
		\hline
	\end{tabular}
\end{table*}

From these distributions of number of companions per star per projected physical separation, we can estimate a binary fraction for each sub-sample of \textcolor{black}{likely} binaries for a specific range of mass ratio and projected physical separation. To do this, we use the log-normal distribution fits from Table \ref{tab:binary_fractions} and integrate over the well-probed projected physical separation range of the sample:
\begin{equation}\label{eq:bin_frac}
	f = \frac{1}{100} \int_{s_{min}}^{s_{max}} A \ e^{\frac{-(log_{10}(s) - \mu)^2}{2 \sigma^2}} ds
\end{equation}
In the above, the $1 / 100$ term is in place to negate the arbitrary scaling applied to the normalization, $s$ is the projected physical separation, and $A$, $\mu$ and $\sigma$ are the model parameters for our log-normal distribution used to describe the projected physical separation distribution for each sub-sample. When estimating binary fractions, we assume the fraction only covers binary mass ratios of $0.6 \lessapprox q \leq 1$  for all sub-samples. This selection seems sound for the \textcolor{black}{K} dwarfs based on the range of $|\Delta G|$ probed \citep{pecaut2013}, but it should be noted that the M dwarfs do probe lower mass ratios than this for $|\Delta G_{max}| = 2$. Additionally, we set $s_{min/max}$ in the range of $s_{min} =30$ AU and $s_{max} = 500$ AU for all sub-samples, as we do not believe we are fully probing the closest separation binaries to a level of statistical significance. Overall, this means that all of our binary fractions will only probe the fraction of binaries in the range of $q=[0.6, 1]$ and $s=[30, 500]$ AU.

The resulting binary fractions for all sub-samples are shown in the last three columns of Table \ref{tab:binary_fractions} for various ranges in physical separation. For these estimates, errors in the binary fraction are found by bootstrapping the fraction for 10,000 iterations based on the posterior distribution for the model parameters for the \textcolor{black}{log-normal} fit. Similar to \citet{Susemiehl2022}, we find that \textcolor{black}{for both spectral types} we get a comparable binary fraction (within the uncertainties).

To more specifically consider the M dwarfs, using the M dwarf binary distribution found in \citet{Susemiehl2022} with eq.~\ref{eq:bin_frac}, we get a binary fraction of $0.268 \pm 0.053$ for the projected physical separation range probed here, which is slightly lower, but comparable, to the value found with our fit (Table \ref{tab:binary_fractions}). Additionally, \citet{Susemiehl2022} found that $10^\mu = 47.863_{17.633}^{15.429}$, which is higher than our fit here, though not to a level of statistical significance. Both the peak in this study and in \citet{Susemiehl2022} are also significantly larger than the broad peak of $\sim 20$ AU found by \citet{winters2019}. For this study, this is most likely due to the primary mass probed for the M dwarfs. For the M dwarf cut, we chose to only select M dwarfs with $8.15 < M_G < 13$, which roughly corresponds to $0.16 < M_\odot < 0.57$ \citep{pecaut2013}, as for this large volume there are very few M dwarfs, relatively, at fainter absolute magnitudes. This seems to explain the difference in the peak between this study and \citet{winters2019}, as \citet{winters2019} found that early-type M dwarfs tended to have fewer binaries at lower separations as compared to late-type M dwarfs.

\textcolor{black}{The above demonstrate that with this sample we are able to find some interesting results in regard to binary science. We do find that with the current sample though, it is difficult to e.g., probe smaller separations or look at various Galactic populations to see how these distributions may change. So, while our catalog is capable of investigating interesting binary science questions on its own, its true power may come from combining it with other catalogs.} For example, metallicity measurements from SDSS-V may \textcolor{black}{help} constrain the binary fraction for various \textcolor{black}{Galactic} populations, or the addition of radial velocity measurements from Gaia DR3 may better allow \textcolor{black}{binaries} to be divided into \textcolor{black}{Galactic} populations via total space motion. For the current study though, this result does show promise in using this catalog to study the physical properties of nearby binaries of varying spectral types.


\section{Conclusions}\label{sec:conclusions}

In this study, we examined the small angular separation neighbors to nearby field stars in Gaia eDR3. We identified significant excesses in 2MASS photometry relative to the Gaia photometry, which is consistent with the presence of an additional star in the field at the epoch of 2MASS. This confirms that most of these neighbors are in fact true visual companions, and not spurious entries in Gaia eDR3. We demonstrate that the observed relationship between $G-J$ excess and $\Delta G$ is consistent with the expected relationship for stars at the same distance (i.e. in a physical binary system). But, we also show that for some of the alleged chance alignments, the $G-J$ excesses are also often consistent with with a binary system.

To better differentiate binaries and chance alignments, we consider a higher dimensional distribution consisting of the $G$ magnitude of the primary, sine of the Galactic latitude, Galactic longitude, $|\Delta G|$, $G-J$ excess\textcolor{black}{, angular separation between the stars and the \texttt{ipd\_frac\_multi\_peak} value from \textit{Gaia}}. The PDF of these distributions \textcolor{black}{were} estimated using a KDE, and by dividing the probability of a pair from the chance alignment distribution by the probability from the candidate distribution, we get \textcolor{black}{a ``contamination factor", which we use to determine how likely} every pair of stars \textcolor{black}{is to being} a chance alignment. As this \textcolor{black}{contamination factor} is not strictly a probability, we calibrate the \textcolor{black}{contamination factor} value \textcolor{black}{assuming the distribution of binaries follows the sky distribution of the 200 pc sample. This expected distribution allows us to determine the contamination and completion rate across the sky and identify an ideal value to select likely binary based on their contamination factor}. By selecting pairs with with \textcolor{black}{$L < 0.00193$}, we are able to \textcolor{black}{get an overall completion rate 75\% with a contamination rate of 14.1\%. By cleaning the sample of likely binaries by removing binaries in high contamination regions on the sky ($C_{ideal} > 0.1$), we can get a sample with a much lower contamination rate of 0.4\%.}

Overall, this results in a catalog of \textcolor{black}{68,725} likely candidate binaries (or \textcolor{black}{50,230} if we exclude the high contamination regions). Less than half of these systems have been previously identified in other Gaia eDR3 wide binary catalogs \citep[i.e.][]{elbadry_2021}. In addition to the large number of newly identified binaries, we demonstrate that our \textcolor{black}{likely} binaries push to smaller angular and physical separations than \citet{elbadry_2021}, allowing for the detection of binaries with shorter projected physical separations. With these \textcolor{black}{likely} binaries we then demonstrate two science cases for such a catalog.

First, we discuss the \textcolor{black}{590} previously unidentified binary systems in Gaia eDR3 with $s<30$ AU that are found in our \textcolor{black}{likely binary} list. This presents a large increase from the 138 within 200 pc previously been identified in Gaia eDR3 \citep{elbadry_2021}. Additionally, we do a literature search to identify systems that have past observations via high resolution imaging or astrometric anomaly, and find that \textcolor{black}{4} out of the \textcolor{black}{15}  shortest separation systems ($s<10$ AU) do not currently have any published observations. In the future, observations of these short period binaries will be crucial to, first, confirm them as binaries and, second, begin to plot out the orbits of these systems. These orbits will allow for mass determination of these stars, which are crucial for the study of stellar parameters.

Lastly, we demonstrate a science case with our catalog of \textcolor{black}{likely} binaries where we estimate projected physical separation distributions for sub-samples of \textcolor{black}{K and early M dwarfs}. From this example we demonstrate that we can well constrain the projected physical separation distribution for the lower mass sub-samples and use these fits to estimate binary fractions for a set range in binary mass ratio and physical separation. We find that these fractions are consistent with trends from previous studies. \textcolor{black}{While this provides an interesting, initial use case of the catalog resulting from this work, we do think the true power of this catalog will come when it is combined with other lists of known binaries from Gaia and other sources. By combing such catalogs, this will allow for} more robust studies of binaries in the Solar Neighborhood where our catalog of \textcolor{black}{likely} binaries provides a needed complement to other, larger catalogs of nearby binaries \textcolor{black}{due to the smaller angular and physical separations, and mass ratios it probes.}

\section*{Acknowledgments}

Mr.~Medan gratefully acknowledges support from a Georgia State University Second Century Initiative (2CI) Fellowship.

This work has made use of data from the European Space Agency (ESA) mission
{\it Gaia} (\url{https://www.cosmos.esa.int/gaia}), processed by the {\it Gaia}
Data Processing and Analysis Consortium (DPAC,
\url{https://www.cosmos.esa.int/web/gaia/dpac/consortium}). Funding for the DPAC
has been provided by national institutions, in particular the institutions
participating in the {\it Gaia} Multilateral Agreement.

This work makes use of data products from the Two Micron All Sky Survey, which is a joint project of the University of Massachusetts and the Infrared Processing and Analysis Center/California Institute of Technology, funded by the National Aeronautics and Space Administration and the National Science Foundation.

%

\bibliographystyle{aasjournal}
\bibliography{GaiaEDR3_Neighbors_ApJ_arxiv}

\end{document}